\documentclass{aa}  
\usepackage{txfonts}
\usepackage{pdflscape}
\usepackage{longtable}

\usepackage{newtxtext,newtxmath}

\usepackage[T1]{fontenc}
\usepackage{ae,aecompl}

\usepackage{graphicx}	
\usepackage{amsmath}	
\usepackage{amssymb}	
\usepackage{rotating}
\usepackage{listings}
\usepackage{threeparttable}

\begin{document} 
             \title{Update of the INTEGRAL/IBIS AGN catalogue: deeper on the Galactic plane and wider beyond}
             
             \subtitle{}

\author{A. Malizia, \inst{1}
L. Bassani,  \inst{1}
R. Landi, \inst{1}
M. Molina, \inst{2}
N. Masetti, \inst{1,3}
E. Palazzi, \inst{1}
G. Bruni, \inst{4}
A. Bazzano, \inst{4}
P. Ubertini, \inst{4}
A.~J. Bird, \inst{5}
}

\institute{OAS-Bologna, Via P. Gobetti 101, 40129 Bologna, Italy\\
              \email{angela.malizia@inaf.it}
         \and
               IASF-Milano, Via Corti 12, 20133 Milano, Italy
         \and
            Instituto de Astrof\'isica, Facultad de Ciencias Exactas, Universidad Andr\'es Bello, Fern\'andez Concha 700, Las Condes, Santiago RM, Chile
            \and
               IAPS-Roma, Via Fosso del Cavaliere 100, 00133, Roma, Italy 
          \and
               School of Physics and Astronomy, University of Southampton, Southampton SO17 1BJ\\
}

\date{ Received ; accepeted}

\abstract
{In this work we have updated the list of  AGN detected by INTEGRAL taking into account the new objects listed in the last published INTEGRAL/IBIS survey. 
We have collected 83 new AGN increasing the number of INTEGRAL detected active galaxies (436) by 19\%. 
Half of these new additions are located behind the Galactic plane; for most of them we have full X-ray coverage obtained through archival data from Swift/XRT, XMM-Newton and NuSTAR. 
This allowed us to associate each high-energy emitter with a single or multiple X-ray counterpart/s and characterise the spectral shape of these new AGN by estimating the photon index, the intrinsic absorption and the 2–10 keV flux. A few cases where two soft X-ray counterparts fall within the INTEGRAL error circle and at least one is classified as an AGN have been found and discussed in detail. 
Thirty-four sources originally listed as AGN candidates or unidentified objects have been recognised as AGN by employing three diagnostic tests: WISE colours, radio emission and morphology. For 12 sources, among the 34 AGN candidates, we reduced optical spectra and confirmed their AGN nature, providing also their optical class and redshift. 
This paper is part of an on-going effort to keep the INTEGRAL AGN catalogue updated in order to provide the scientific community with a hard X-ray selected sample of active galaxies well classified and spectrally characterised.}

  \keywords{X-rays: galaxies -- galaxies: active -- galaxies: Seyfert }

\titlerunning{Update of the INTEGRAL/IBIS AGN catalogue}
\authorrunning{A. Malizia et al.}
\maketitle



\section{Introduction}
In the last decades the science of Active Galactic Nuclei (AGN) has made great progress which has led to a deep knowledge of both their population in the sky and the physical mechanisms responsible for their emission over the entire electromagnetic spectrum. 
One of the most important results recently proven is the connection between the super massive black hole (SMBH) which powers the AGN, and the host galaxy, the so-called feeding and feedback cycle of active galaxies. 
But, although the co-evolution AGN-host galaxy scenario is widely accepted and supported by clear observational evidence \citep{2013ARA&A..51..511K,2014ARA&A..52..589H}, there are still some issues 
to be addressed mostly regarding AGN feedback  (e.g. \citealt{2012ARA&A..50..455F}).
What is certain as demonstrated by deep X-ray observations, is that most SMBHs 
at galaxy centres grow hidden by dust and gas (see e.g. \citealt{2018ARA&A..56..625H} for a review).

Meanwhile at small scales closer to the SMBH, the standard simplistic picture of the AGN unified model \citep{1993ARA&A..31..473A} is now outdated.
According to this theory, most of the amount of obscuring material mainly resides in the torus which is also solely responsible for AGN classification.  
Actually the present view of AGN is much more complex.
Absorption has been revealed to be on different scales/structures, winds, outflows, jets and of different nature, neutral or/and ionised (see \citet{2017NatAs...1..679R} for a review); 
the absorbing torus itself has a structure not yet well understood \citep{2021A&A...650A..57Z}, and nor is its connection to its surroundings (e.g. \citealt{2015ARA&A..53..365N}).

Therefore absorption is proven to be a fundamental key both for the understanding of the geometry of the innermost part of the AGN and for feedback with the host galaxy.
The most efficient way to study absorption and address these still open issues, is to have large samples of AGN selected in the hard X-ray band ($>$ 20 keV) which is far less biased towards absorption.
Since the early 2000s both INTEGRAL/IBIS \citep{2003A&A...411L.131U} and Swift/BAT \citep{2005SSRv..120..143B}, thanks to their good sensitivity and wide-field sky coverage in the high-energy domain (14-200 keV), have provided a remarkable improvement in our knowledge 
of the high-energy extragalactic sky by detecting more than 2000 (mostly local) AGN. 
It is worth noting that, due to its observational strategy, Swift/BAT performs a nearly uniform all-sky survey and so it is more effective at higher Galactic latitudes. INTEGRAL, on the other hand, provides a sky survey with exposures that are deeper in the Galactic plane and Galactic centre regions. INTEGRAL has also 
a higher angular resolution, which is essential in these crowded regions.
This makes the two observatories fully complementary also in the case of extragalactic studies; INTEGRAL surveys were particularly important in detecting  highly absorbed AGN along the Galactic plane.
The area comprised between $\pm$10-15 degrees above and below the Galactic plane, the so-called "Zone of Avoidance", is rich of
gas and dust which obscure starlight  and screen nearly all background extragalactic objects from traditional optical-wavelength surveys (in the optical, as much as 20\% of the extragalactic sky is obscured by the Galaxy). 
As a consequence, historically the Galactic plane has not been a focus for extragalactic astronomy before the advent of INTEGRAL. Hard X-rays ($>$10 keV) are able to penetrate this zone providing a window that is virtually free of obscuration relative to optical wavelengths and partly also to soft X-rays.
The capabilities of INTEGRAL/IBIS in studying extragalactic sources were revealed soon after the launch of the satellite.
It became soon evident that the population of AGN emitting above 20 keV was growing thanks to the discovery that many of the new hard X-ray detected sources (IGR sources) were indeed active galaxies. 
In the first IBIS survey \citep{2004ApJ...607L..33B}, based on the first year of INTEGRAL observations, there were listed only 5 AGN which became 33 (almost 20\% of the entire catalogue) in the second survey \citep{2006ApJ...636..765B}.
Since 2004, a sequence of IBIS all-sky survey catalogues (\citealt{2016ApJS..223...15B} and references therein) as well as deep surveys of the Galactic centre region and Galactic plane \citep{2004AstL...30..382R, 2006AstL...32..145R, 2007A&A...475..775K, 2012A&A...545A..27K, 2017MNRAS.470..512K} have been published at regular intervals, making use of an ever 
increasing dataset. 
As the total number of known or newly discovered AGN grew, it was possible to assemble them for population studies as we have done over the years 
\citep{2012MNRAS.426.1750M, 2016MNRAS.460...19M, 2020A&A...639A...5M}. 
So far the number of AGN listed in this dataset amounts to 436 objects.

Taking advantage of the hard X-ray selection, this large sample of AGN has been exploited to carry on several studies on relevant issues of extragalactic science such as the possibility to identify highly obscured AGN like Compton thick sources more easily \citep{2007ApJ...668...81M}. 
It also allowed to tighten the constraints on the absorption properties of the local AGN
population, finding that the true fraction of  Compton thick AGN is $\sim$ 20\% 
\citep{2009MNRAS.399..944M}. Furthermore, broad-band spectral (0.2--100 keV ) analysis of a complete sample of Seyfert 1 galaxies, allowed \citet{2014ApJ...782L..25M} to constrain for the first time the value of the high-energy cut-off at around 100 keV, clearly indicating that the primary continuum typically decays at much lower energies than previously thought. 
Alongside INTEGRAL, Swift/BAT has also contributed to these issues mostly confirming INTEGRAL results \citep{2011ApJ...728...58B,2015ApJ...815L..13R, 2017ApJS..233...17R,2022ApJ...927...42K}.

Given the scientific relevance of the above results, it is fundamental to keep updating these AGN high energy catalogues.
The last INTEGRAL all-sky survey by \citet{2022MNRAS.510.4796K} collects all the INTEGRAL observations over 17 years until January 2020, 
and lists 929 hard X-ray sources detected above the 4.5$\sigma$ threshold, of which 113 are reported as unclassified.
In order to update the catalogue of INTEGRAL AGN, in this work we have added from Krivonos' list the AGN which have not been reported as INTEGRAL detection before; we have also searched for their soft X-ray counterparts in order to study their spectral behaviour.
Moreover, we have tried to pinpoint, among unidentified sources, those likely to be AGN on the basis of diagnostic indicators such as WISE colours, radio emission and indication of extension in the optical images, we have also classified 12 among AGN candidates by analysing their  optical spectra.

\section{New AGN, the sample}
\begin{figure}
	\includegraphics[width=8cm]{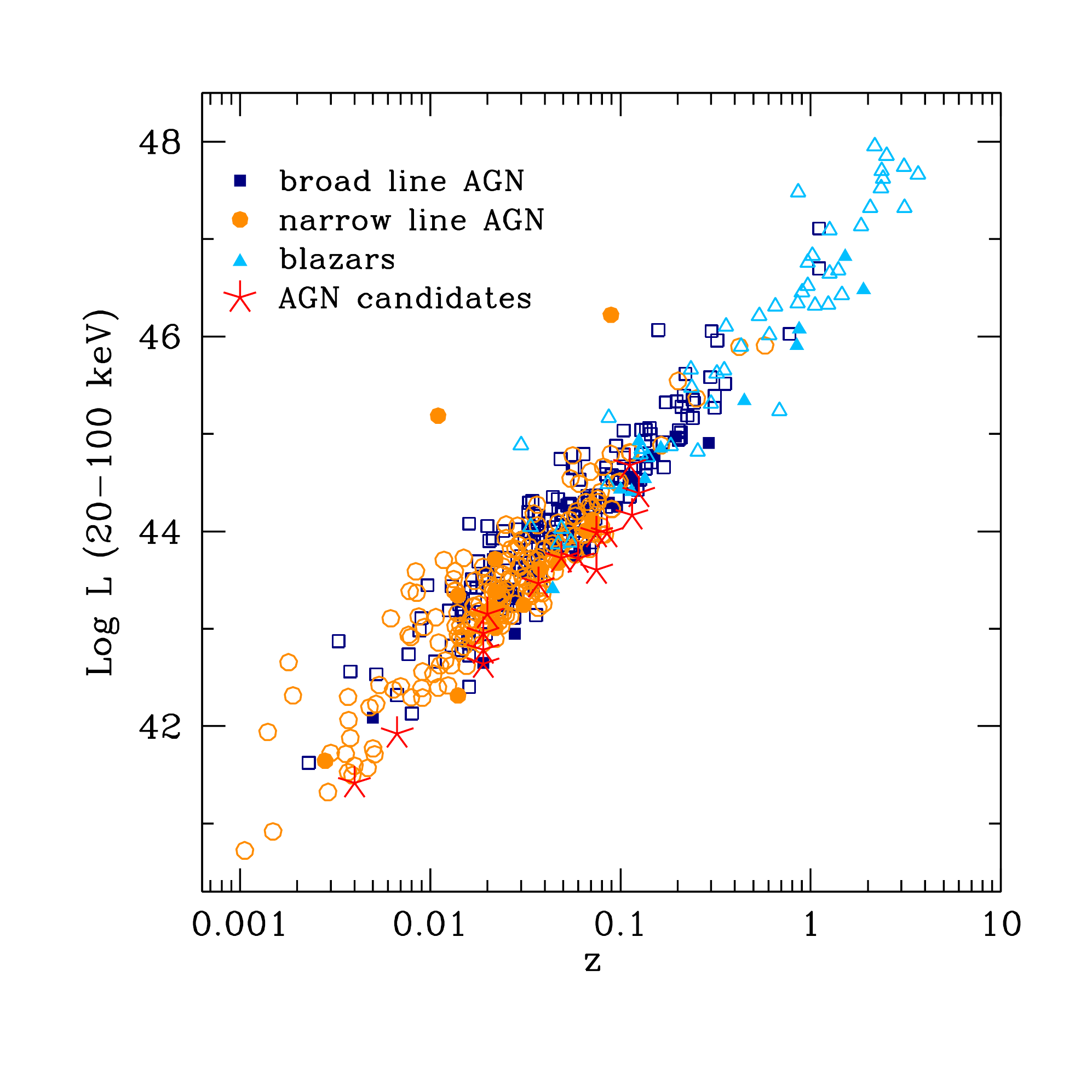}
\caption{Observed hard X-ray (20-100 keV) luminosity versus redshift for the whole INTEGRAL AGN sample. Blue squares are broad line AGN, gold circles are narrow line AGN,  and light blue triangles are blazars. Red stars are AGN candidates still unclassified for which a measure of redshift is available. The filled symbols referred to the new AGN added in this work.}
    \label{fig1}
\end{figure}

\begin{figure*}
\includegraphics[width=\textwidth]{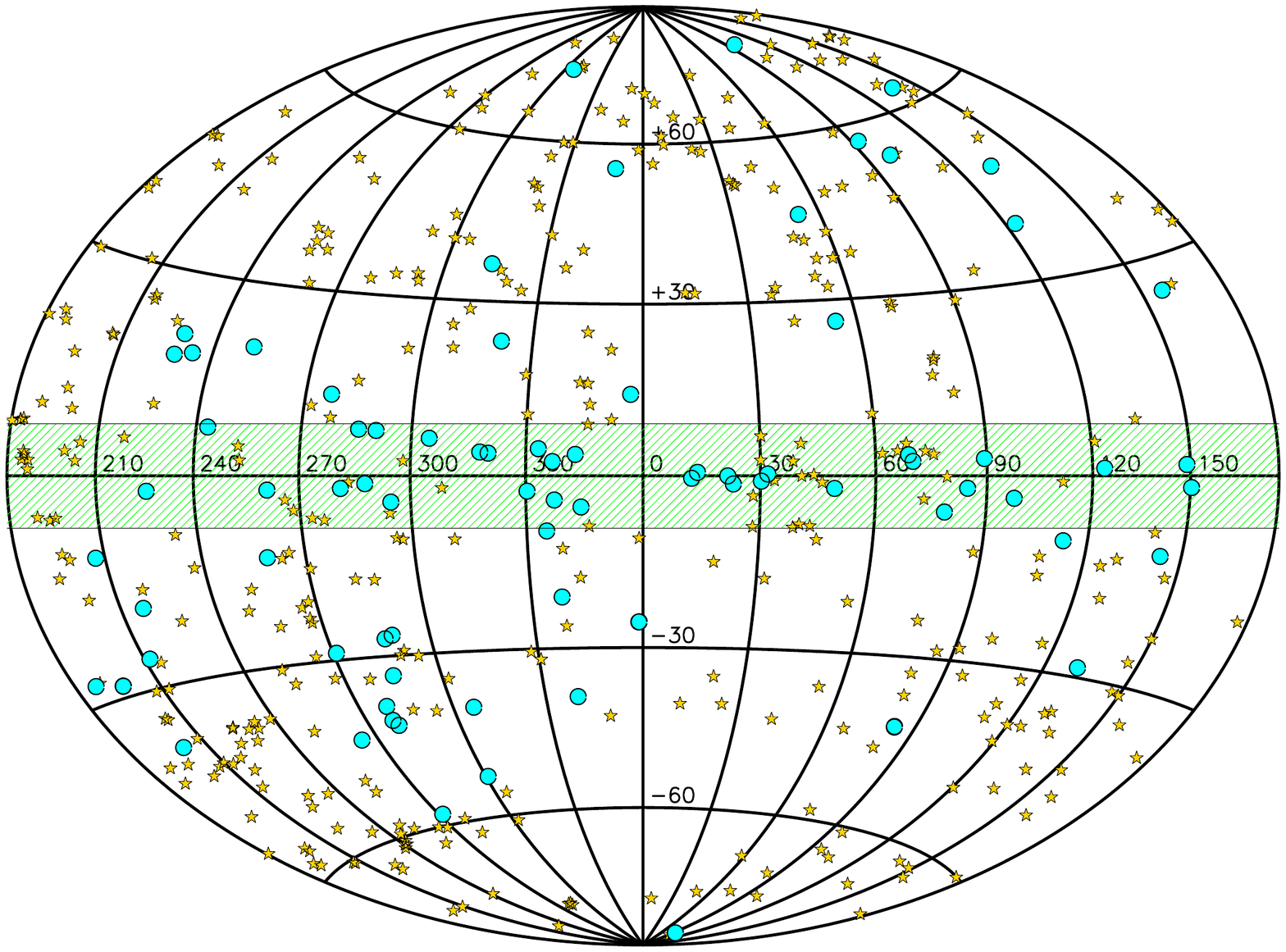}
\caption{All the AGN detected so far by \emph{INTEGRAL/IBIS} plotted on the sky. The circle represent the 83 new active galaxies studied in this work, while the stars are the AGN detected in previous surveys and reported in \citet{2012MNRAS.426.1750M, 2016MNRAS.460...19M, 2020A&A...639A...5M}}
\label{fig2}
\end{figure*}

The aim of this work is to update the list of  AGN detected by INTEGRAL taking into account the new 
active galaxies already listed in  the \citet{2022MNRAS.510.4796K} survey. 
We have collected 83 new AGN listed in Table \ref{tab1}, among these 4 sources have already been listed in \citet{2016MNRAS.460...19M} as AGN candidates and reanalysed here in order to be optically classified; 3 objects, highlighted in bold in the table, are from  \citet{2016ApJS..223...15B}; 34 objects are unclassified AGN, of which 22 have been already proposed as AGN candidates, while the rest, 12, are  unidentified objects in the \citet{2022MNRAS.510.4796K} survey and have been classified or proposed as AGN in the present work (see section 4).
In Fig. \ref{fig1} the entire INTEGRAL AGN sample so far collected  is shown with the different classes differentiated (see figure caption). The last additions have been highlighted with filled symbols.
As clearly seen from Fig. \ref{fig1} the new AGN collected in this work (filled symbols) well reside 
in the parameter space of the INTEGRAL AGN sample (see AGN review by \citealt{2020NewAR..9001545M}), but what it is worthy of note is that a large fraction of them are located on the Galactic plane.
In Fig. \ref{fig2} the entire sample of INTEGRAL AGN are plotted in the sky; the stars are the AGN reported in previous catalogues \citep{2020A&A...639A...5M,2016MNRAS.460...19M,2012MNRAS.426.1750M}, while the circles are the 83 AGN collected in this work.
It is evident that a large number of the new additions are located in the plane of the Galaxy: 43\% of the sample are within  $\pm$10 degrees above and below the Galactic plane; the percentage rises to almost half of the sample ($\sim$50\%) within $\pm$15 degrees.

We have been able to obtain full X-ray coverage of almost the entire sample by making use of data from Swift/XRT, XMM-Newton and NuSTAR archives. Only for two objects, IGR J06380-7536 and RX J1317.0+3735, there are no 2-10 keV data available but only a soft X-ray flux from the XMM-Slew Survey \citep{2008A&A...480..611S}. 
2-10 keV data have been used to associate the high-energy emitter with a single or multiple X-ray counterpart/s, getting a better (arcsec) position. For the counterpart search we have adopted the positional uncertainties quoted by \citet{2007A&A...475..775K} (see next section); hard/bright  X-ray sources have been favoured over soft/dim objects.
The soft X-ray data have also been used  to characterise the high-energy source at low energies.
Unfortunately in most cases the statistically quality of the 2--10 keV data is poor and therefore we have employed a simple model consisting of an 
absorbed power law which allowed us to estimate the main spectral parameters such as the intrinsic absorption\footnote{the Galactic absorption has been always considered} and the 2–10 keV flux (e.g. \citealt{2016MNRAS.460...19M}).

In Table \ref{tab1} we have listed the new AGN together with their essential information as always reported in our catalogues: alternative names (column 2), the most precise coordinates RA and Dec (columns 3 and 4) from optical or soft X-rays (Chandra or Swift/XRT), redshift (column 5), optical class taken from NED, Simbad archives or \citet{2010A&A...518A..10V} (column 6). The  X-ray spectral parameters reported in this work are: column density, photon index and 2-10 keV flux (columns 7 - 9); the  hard X-ray flux (20-100 keV, column 9) has been extrapolated from \citet{2022MNRAS.510.4796K}. In column 11 the reference for the 2-10 keV spectral parameters is given if not provided by the data analysis performed for the first time in the present work.

\section{Cases with multiple counterparts}
In this new sample of AGN detected by INTEGRAL we have found a few cases where two soft X-ray counterparts fall within the INTEGRAL error circle and at least one is classified as AGN. These objects are discussed individually in the following sub-sections.
In this work, following \citet{2007A&A...475..775K}, we have used a 4.2 arcmin search radius (90\% confidence level for a source detected at 5 - 6 standard deviations) or, in case of no detection of a counterpart, a 6.5 arcmin (99\%), to search for possible X-ray counterparts. Only counterparts coincident with well defined active galaxies are reported in Table \ref{tab1}, while other potential associations are briefly described in this section.
In the Appendix we show the Swift/XRT 0.3-10 keV images of sources discussed below.
It is worth noting that in those cases where both sources are detected above 3 keV, the hard X-ray emission detected by INTEGRAL and reported in \citet{2022MNRAS.510.4796K} has to be considered as the combination of the two sources, therefore in column 9 of 
Table \ref{tab1} only estimates of 20-100 keV fluxes have been provided.

\subsection{IGR J04288-6702 also SWIFT J0428.2-6704 A/B} 
In the case of IGR J04288-6702 (see Fig. \ref{a1}, left panel) we find two objects emitting above 3 keV which are both outside the 90\% (filled circle) but inside the 99\% (dashed circle) IBIS positional uncertainty: one (\#1) is the Seyfert 1.5 galaxy LEDA 299570 reported in the Table \ref{tab1} and also reported in \citet{2022MNRAS.510.4796K}, while the other one (\# 2) at $\sim$11.5 arcmin, is the Galactic source SXPS J042749.2-670434 (RA(J2000) = 04 27 49.63, Dec(J2000)= -67 04 34.8), also reported as a Fermi/LAT source, 4FGL J0427.8-6704 (see dashed-dotted error ellipse in Figure \ref{a1}, left panel).
The latter object is associated with an eclipsing low mass X-ray binary (P = 8.8 hr) with a main-sequence donor and a neutron-star accretor \citep{2016ApJ...831...89S,2020A&A...638A.128M}.  
The X-ray light curve also shows properties similar to those seen  among known transitional millisecond pulsars: short-term variability, a hard power-law spectrum ($\Gamma$ $\sim$ 1.7), and a comparable 0.5-10 keV luminosity (2.4 $\times$10$^{33}$ erg s$^{-1}$). The source is furthermore remarkable for its face-on inclination in the range 5$^{\circ}$-8$^{\circ}$, depending on the neutron star and donor mass \citep{2017ApJ...849...21B}. Moreover, a gamma-ray eclipse having the same phase as the optical one has been recently reported together with an X-ray eclipse \citep{2020MNRAS.494.3912K}. This is clearly a very interesting object which deserves further studies especially in the hard X-ray band. NuSTAR, having an angular resolution better than IBIS, can allow us to discriminate and study both these two  sources.

\subsection{PKS 0440-00 and 1RXS J044229.8-001823} 
In the right panel of Fig. \ref{a1} we show a very peculiar case as two blazars fall within the IBIS error circle: PKS J044-00 also known as QSO B0440-003 (\#1), a flat spectrum radio quasar at z=0.845 and 1RXS J044229.8-001823, also known as 2MASS J04423023-0018294 (\#2), a BL Lac object at z=0.449.  
For this reason it is incorrect to quote PKS 0044-00 as the only INTEGRAL detection as listed in \citet{2022MNRAS.510.4796K} survey catalogue; 
we suggest instead that this source should be called IGR J04426-0018, since its emission  is the merger of two objects.
The QSO is also detected up to the GeV band by Fermi/LAT  and reported as 4FGL J0442.6-0017 (\citealt {2020ApJ...892..105A}, see dashed-dotted error ellipse in the figure) in the fourth AGN  catalogue, while the BL Lac, originally indicated in the third Fermi catalogue as possible counterpart, has been excluded in latest one. However, the BL Lac object is listed in the 3HSP catalogue of high synchrotron-peaked blazars \citep{2019A&A...632A..77C}, showing a synchrotron peak at around 2 $\times$10$^{16}$ Hz. To estimate the 20-100 keV flux of each source, we assumed an equal share of emission as implied by the almost equal flux at soft X-ray energies and a similar spectral shape at X-ray energies. Also this would be an interesting sky field to be pointed by the NuSTAR telescope.

\subsection{SWIFT J0835.5-0902} 
As evident in the XRT image shown in Fig. \ref{a2} (left panel), here too we find two hard X-ray emitters associated with a single INTEGRAL/BAT source, although a unique identification is provided in the 157-month Swift/BAT catalogue \citep{2018ApJS..235....4O}. The first object (\#1), the one reported in Table \ref{tab1}, is located at RA(J2000) = 08 35 33.34 and Dec(J2000) = 09 05 30.2 (3.5 arcsec X-ray positional uncertainty) and  associated with the extended infrared object 2MASX J08353333-0905302. This galaxy is reported in Simbad with no redshift, while NED provides a photometric redshift of z=0.1124. The WISE colours, W1-W2 = 0.86 and W2-W3 = 2.87, are typical of AGN \citep{2015ApJS..221...12S}; the source is listed in the UV-bright QSO survey of 
\citet{2016AJ....152...25M} and it is also identified with an active galaxy from multi-waveband analysis \citep{2012ApJ...751...52E}. The lack of intrinsic absorption in the X-ray spectrum indicates that this is most likely a broad line AGN. 

The other counterpart (\#2), located at RA(J2000) = 08 35 31.5, Dec(J2000) = -09 04 18.6 (3.8 arcsec X-ray positional uncertainty), is a Galactic source. It is also reported as an XMM Slew source with 0.2-12 keV flux in the range 2.4-3.6 $\times$10$^{-12}$ erg cm$^{-2}$ sec$^{-1}$. The XRT spectrum is well fitted by a power law with photon index of 1.24$\pm0.27$ and a 2-10/0.2-12 keV flux of 8.1$\times$10$^{-13}$/1.2$\times$10$^{-12}$ erg cm$^{-2}$ s$^{-1}$, lower than that measured by XMM-Newton. The source is associated with the variable star ASAS J083531-0904.1\footnote{\url{https://www.aavso.org/vsx/index.phpview=detail.top&oid=281217}}, classified as a rotating variable at a distance of 4.3 Kpc (GAIA parallax) and with a possible period of 171.2 day. However, it is worth noting that in the GAIA colours versus absolute magnitude diagram (BP-RP=2.1, absolute magnitude = -1.83) the source is located in the region populated by eclipsing binaries. It would be interesting to study this source further in order to unveil its true nature and its contribution to the overall IBIS emission.

\subsection{IGR J12418+7805}   
Another interesting case is IGR J12418+7805, where possibly two objects contribute to the overall emission seen by INTEGRAL but, unfortunately, no Swift/XRT observation is available.
We found one source lying within the IBIS error circle, which is clearly reported as an X-ray emitter both in the ROSAT Bright Source \citep{1996IAUC.6420....2V} and in the XMM Slew Survey \citep{2008A&A...480..611S}, and is therefore listed in Table \ref{tab1}. The source is associated with the galaxy LEDA 140000 at a redshift of z=0.022, which is classified as a Seyfert 1.9 in the \citet{2010A&A...518A..10V} catalogue and has a 0.2-12 keV flux of  2.1 $\times$10$^{-12}$ erg cm$^{-2}$ sec$^{-1}$.
Assuming a typical value 1.8 for the power law photon index, we have extrapolated the 2-10 keV flux value reported in Table \ref{tab1}. 

The other source, reported only in the XMM Slew Survey catalogue, is at $\sim$ 2.2 arcmin distance from the previous one and is  located at RA(J2000) = 12 43 13.8, Dec(J2000) = +78 08 28, with an associated positional uncertainty of $\sim$10 arcsec. It is brighter than LEDA140000 with a 0.2-12 keV flux of 6 $\times$ 10$^{-12}$ erg cm$^{-2}$ sec$^{-1}$; the source class is unclear, although its location at high-galactic latitude suggests an extragalactic nature. 
The two XMM Slew sources were detected at different epochs, almost 1 year apart, and no radio, infrared nor optical association could be found within the XMM Slew positional uncertainty of the second source. Only enlarging its positional uncertainty to 15 arcsec, it is possible to find a WISE source with W1-W2=0.47 and W2-W3=4.04, which are more typical of starburst galaxies than AGN \citep{2017ApJ...836..182J}. Due to these uncertainties, we decided not to list this second source among the current INTEGRAL AGN update, waiting for a more sensitive hard X-ray image of this sky region to confirm its contribution to the overall INTEGRAL emission.

\subsection{SWIFT J1937.5-4021} 
SWIFT J1937.5-4021 is reported as an unidentified object in the \citet{2022MNRAS.510.4796K} list but we found that the most likely soft X-ray counterpart for both IBIS and BAT detection is the XRT source located at RA(J2000) = 19 37 13.47 and Dec(J2000) = -40 16 14.84 (3.6 arcsec error radius), detected at 27$\sigma$ in the 0.3-10 keV band (20$\sigma$ above 3 keV) and well inside the IBIS error circle (see source \#1 in Fig. \ref{a2}, left panel).
This X-ray source is also reported as an XMM Slew source with a 0.2-12 keV flux of 1.37 $\times$10$^{-12}$ erg cm$^{-2}$ s$^{-1}$  and associated with the galaxy LEDA 588288 with z=0.075 
in the Simbad archive but reported in NED with a photometric redshift of z=0.055776.  
Our analysis of the X-shooter optical spectrum of this source allowed us to confirm its nature, classifying it as  Seyfert 2 galaxy, and measure its true redshift of z=0.0193 (see section 4.2).

The other possible counterpart, although outside the 90\%  but inside the 99\% IBIS error circle, is associated with the ROSAT source 1RXS J193716.1-401026, located at RA(J2000) = 19 37 14.70 and Dec(J2000) = -40 10 14.91 (3.64 arcsec positional uncertainty).  
Swift/XRT detected this source at 23$\sigma$ in the 0.3-10 keV band (9$\sigma$ above 3 keV). 
As evident from Fig. \ref{a2} (right panel) both sources are well inside the 90\% BAT error circle (dotted circle). 
We suggest for 1RXS J193716.1-401026 an extragalactic origin since GAIA does not detect a significant parallax or proper motion \citep{2019RAA....19...29L}; moreover, the observed WISE colours (see section 4.1) are typical of AGN.
Both XRT detections are hard X-ray sources, confirming that both can contribute to the high-energy emission seen by IBIS.

\subsection{AX J2254.3+1146  also  SWIFT J2254.2+1147 A/B}  
This is another case where two AGN fall within the IBIS positional uncertainty, although one is well inside the error circle while the other is just at its border but inside the 99\% error circle (see Fig. \ref{a3}). Both sources are well detected above 3 keV, they have similar redshifts but no interaction is evident between the two. 
The first object (\#1), UGC 12237, is a Seyfert 2 galaxy, while the other (\#2) is a Seyfert 1 (UGC 12243) according to \citet{2010A&A...518A..10V};
both AGN are hosted in edge on galaxies of 0.24 and 0.34 axial ratio (major over minor axis), respectively. This is also an interesting set of sources to be observed with the NuSTAR observatory in order to study both AGN as already done for IGR J16058-7253 \citep{2021MNRAS.507.3423M}.

\section{Unclassified objects}
Thirty-four sources ($\sim$41\% of the total sample) listed in Table \ref{tab1} are unclassified AGN; 22 of them were already identified as active galaxies in the \citet{2022MNRAS.510.4796K} survey, while the remaining 12 are reported as unidentified sources and 
proposed as AGN candidates in this work. For this set of 34 sources we have investigated their AGN nature by means of a two-step approach. 
First  we used multi-waveband information to asses their nature. In particular, we made use of 3 diagnostics to confirm their nuclear activity: WISE colours, detection of radio emission and other extragalactic features such as location on the sky, measured redshift and  optical morphology, i.e. we have searched for indication of extension in their optical/near infrared images.

As a second step, we searched public  archives for optical spectra of these unclassified sources, finding that 40\% of the sample had indeed public data.
The results of this two-step approach are presented in the following sections.

\subsection{Multiwaveband AGN signature: WISE colours, radio emission and other features}
\begin{figure}
\includegraphics[width=\textwidth,width=8cm]{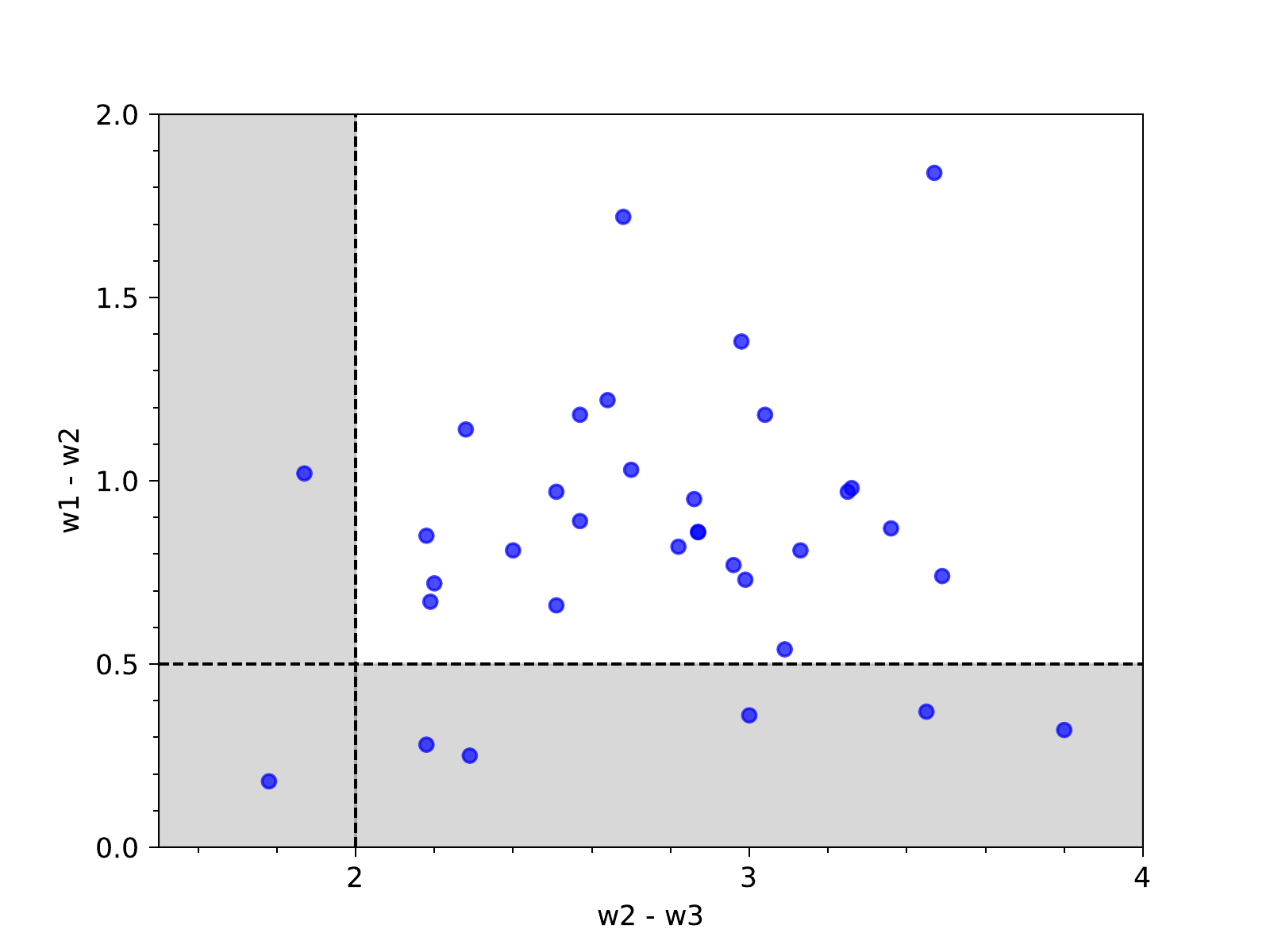}
\caption{WISE colour/colour diagram of the 34 AGN candidates. The  lines from \citet{2015ApJS..221...12S} is to delimit the area populated by AGN. Six sources (IGR J03574-6602, IGR J06075-6148, IGR J06503-7742, IGR J16246-4556, IGR J16459-2325 and IGR J17255-4509) have low W1-W2 values and need to be investigated more deeply. The only source with W2-W3 below 2 is AX J 1830.6-1002, a very peculiar case.}
\label{fig3}
\end{figure}

It is now well established that WISE colours W1-W2 and W2-W3, can be used to select AGN, or more specifically, efficiently accreting  objects. 
AGN are typically located in a well-defined region of the WISE   colour-colour diagram, i.e. in a region  limited by  W1 $-$ W2 > 0.5 and  2 < W2 $-$ W3 < 5.1 (see \citealt{2015ApJS..221...12S} for more details on the boundary delimitation).  
We have therefore collected WISE colours for all the unclassified objects in order to confirm their extragalactic nature. 
For this task we have used the ALLWISE catalogue \citep{2014ApJ...792...30M}  which contains accurate positions, apparent motion measurements, four-band fluxes and flux variability statistics for over 747 million objects detected on the co-added Atlas Images.
For only one source (IGR J19577+3339)  not reported in the ALLWISE  database, we used 
the CatWISE2020 Catalogue \citep{2021ApJS..253....8M}, which however lists only W1 and W2 magnitudes and so provides only one infrared colour. Table \ref{tab2}  lists  the WISE colours of all unclassified sources while Fig. \ref{fig3} shows the corresponding positions in the WISE colour-colour diagram.

Besides the WISE colours, indication for the presence of an active nucleus also comes from radio emission since almost all AGN detected so far by INTEGRAL have a radio counterpart, which is not necessarily radio loud but can emit at a few mJy level. We have therefore investigated the radio properties of unidentified objects using  as a first step old surveys radio catalogues, more specifically, at 1.4 GHz, the  National Radio Astronomy Observatories Very Large Array Sky Survey  (NVSS, \citealt{1998AJ....115.1693C}) and the 
Multi-Array Galactic Plane Imaging Survey (MAGPIS, \citealt{2006AJ....131.2525H}) while at 0.83 GHz  the Molonglo Telescope  Surveys (SUMSS, \citealt{2003MNRAS.342.1117M}) and MGPS-2  \citep{2007MNRAS.382..382M}. 
These  surveys  are similar in sensitivity and spatial resolution and cover together the whole sky, making  them  particularly well suited for searching  radio emission from  our unclassified objects.
However, their sensitivity threshold  is around the  10 mJy  level, so that a number of sources dimmer than this flux could not be detected. In these cases, we consulted more recent all sky surveys like the VLA Sky Survery (VLASS, \citealt{2021ApJS..255...30G}) at 3 GHz and the Rapid ASKAP Continuum Survey (RACS, \citealt{2020PASA...37...48M}) at 0.88 GHz.
 The radio fluxes obtained from all  these surveys  are listed in Table \ref{tab2},  which  also specifies  the database used.
 Inspection of Table \ref{tab2}  indicates that most objects are radio emitters, although the majority have  relatively weak fluxes,  confirming in any case the presence of an active nucleus at their centres. At the resolution of the surveys employed, most of these detections correspond to a compact/single component morphology, except for 4 objects which are displayed in Fig. \ref{r1} and Fig. \ref{r2} in the Appendix, which show a more complex radio shape. 
 
 Finally, we adopted other extra-galactic indicators such as the knowledge of the source redshift, the location above or below the Galactic plane or the presence of extended emission typical of a galaxy morphology.
 The evidence of source extension has been obtained from inspection  of  images provided by the Dark Energy Camera optical and near-infrared survey
 (DECaPS, \citealt{2018ApJS..234...39S}) which, surveying the  Galactic plane, is the most suitable for our sources. A note  related to the presence  of a redshift or of extension in optical images as well as an indication of the source location  with  respect the Galactic plane is reported in the last column of Table \ref{tab2}.
 
In order to validate our AGN selection and definitely classify a source as an AGN, we have adopted the criterion that at least 2 of the 3 diagnostics used have to be satisfied.

As clearly shown in Fig. \ref{fig3}, WISE colour values point to an AGN nature for the majority of the objects except for 7 of them, which are characterised by W1-W2 colours below  the 0.5 line   and W2-W3 below  the 2 threshold as adopted by \citet{2015ApJS..221...12S}.
Five of  the 6  sources with values below W1-W2 colour threshold (IGR J03574-6602, IGR J06075-6148,  IGR J06503-7742, IGR J16459-2325 and IGR J17255-4509) are however associated to galaxies with a measured redshift while the only exception, IGR J16246-4556, is  extended in the  DECaPS image. 
On the other hand, the only source  above the  W1-W2=0.5 line and outside (although only marginally so)  the W2-W3=2 boundary,  is AX J1830.6-1002 which is however a very puzzling case.
As pointed out by \citet{2009MNRAS.395L...1B}, AX J1830.6-1002 is an AGN located in the Galactic plane, inside a diffuse radio supernova remnant. It is  spatially coincident with a compact radio source having the X-ray  spectrum typical of a Compton thick  AGN.  Given the complexity of its local environment and considering the uncertainties associated to the WISE magnitudes, we do not consider this source as an  outlier.
Furthermore, it is worth noting that all these 7 sources have radio detection with the exception of IGR J16459-2325 for which only a RACS upper limit is available.

According to \citet{2015ApJS..221...12S} the number of  contaminating stellar objects in their mid-IR AGN sample is very small, below 0.04\%, making their selection   extremely reliable. Moreover, they quote their sample completeness  to be  $\sim$84\%, meaning that 16\% of AGN are not selected using their method mainly because they display a  W1-W2 colour bluer than allowed, i.e. below the 0.5 line. Interestingly this is very close to the percentage of AGN  (18\%) we find below W1-W2=0.5 both in the present sample or considering the entire set of INTEGRAL AGN,  where 507 out of 519 objects are listed in the ALLWISE catalogue.

These unusual  WISE colours  are generally associated to  heavily absorbed AGN as pointed out by \citet{2015MNRAS.449.1845G} or to  low luminosity AGN, for example radiatively inefficient accretors, 
like LINERs  as indicated  by \citep{2016MNRAS.462.2631M}. In objects classified as LINERs the host galaxy dominates the mid-IR emission with respect to the AGN moving them below the 0.5 threshold in  the W1-W2 versus W2-W3 diagram \citep{2016MNRAS.462.1826H}\footnote{Of the 29 Compton thick AGN and  19 LINERs (or Seyfert/LINERs) present in entire AGN  sample, 12 (or 41\%) of the first and 10 (or 53\%)  of the second are located below the W1-W2=0.5 line, suggesting  that these two types of active nuclei  have a far greater chance than other AGN types to fall below the adopted threshold.}.

As can be seen from Table \ref{tab1}, the X-ray  column densities of our 6 outliers indicate that  4 out of 6 objects are indeed absorbed with N$_{\rm H}$ in excess of 10$^{23}$ cm$^{-2}$, suggesting that the first hypothesis is a viable  explanation for their unusual WISE colours. However, as discussed  in Sect. 4.2, the low luminosity AGN possibility turns out to be an equally valid argument.

As evident from values reported in Table \ref{tab2}, all our AGN candidates have radio detections apart for 5 sources (IGR J04085-6546, IGR J14417-5533, IGR J16459-2325, SWIFT J1839.1-5717 and SWIFT J1937.5-4021(2) which have not been detected in the radio band and for which we were able to extract only upper limits from RACS images. 
Nevertheless, for all these 5 sources, WISE colours and morphology features point to an extragalactic nature and, 
therefore, we are reasonably confident that our final AGN selection is correct (see also following section).

It is worth noting that radio data analysis performed in this work was crucial in unveiling the AGN nature of several objects as in the case of  IGR J19577+3339.
This source is the only one with just one WISE colour and  no other diagnostic indicator of being extragalactic.
IGR J19577+3339 is instead  a strong radio emitter (see Fig. \ref{r1}, left panel in the Appendix) which is used in the literature as a  flat-spectrum radio quasar calibrator \citep{2020A&A...644A.159C}; a central bright component is visible in VLASS, with some additional emission at about 5 arcsec towards NE and SW, most probably not correlated with the source. Previous VLBI studies at 2.3 and 8.6 GHz \citep{2012A&A...544A..34P} showed a bent jet on the pc-scale.

Another source for which the radio detection and radio morphology is decisive to assess its AGN nature is IGR J08297-4250. As shown in Fig. \ref{r1} (right panel) in the Appendix, it shows a core-jet like structure in RACS, reminiscent of a blazar source although at a relatively low radio flux. 

Finally, we note that two other objects, namely PKS 1413-36 and IGR J16005-4645, show strong radio fluxes above the 1 Jy level; the first is associated with a radio galaxy of large size (around 430 kpc, see Fig. \ref{r2} left panel), while the second one is so far still morphologically unclassified (although it is reported as extended in some surveys like ATCA, 
\citealt{2012MNRAS.422.1527M}) and poorly studied at radio frequencies. 
The RACS image of IGR J16005-4645 (see Fig.\ref{r2}, right panel) shows a core+jet like structure, but, combining data from various surveys, the radio spectrum is steep (-0.7/-1) over the entire 0.150-20 GHz range. 
This suggests a large viewing angle and a possible radio galaxy morphology with a lobe pointing towards us.
We also note that the radio and X-ray positions are displaced by roughly 28 arcsec, this could be interpreted as a radio galaxy where the radio position refers to one of the two lobes while the X-ray position coincides with the source core. 
Alternatively, one must assume that the radio object is instead composed of two sources not fully resolved in the radio image and associated with different optical counterparts. Indeed, the radio central position is compatible within errors with one weak GAIA EDR3 catalogue source (2020 VizieR On-line Data Catalog: I/350) of unknown class.
In any case, the X-ray source, actually the secure counterpart of the INTEGRAL emitter, is likely an AGN not only on the basis of its WISE colours and possible radio emission, but also according to the SIX index developed by \citet{2012ApJ...751...52E}. This index was empirically constructed by combing infrared colours over the 2-10 micron waveband (to ensure an AGN like continuum) and the distance to the nearest X-ray source.
A value below zero of the resulting SIX index (-0.258 in our case) indicates an object with a high likelihood (greater than  95$\%$) of being an AGN.

\subsection{Optical spectroscopy}
For 12 sources, among the sample of 34 AGN candidates, we acquired reduced optical spectra available in the 6dF\footnote{available at: \url{http://www-wfau.roe.ac.uk/6dFGS/form.html.}}, 
PESSTO\footnote{available at: \url{https://www.eso.org/qi/catalogQuery/index/365.}} 
and X-Shooter\footnote{available at: \url{http://archive.eso.org/eso/eso_archive_main.html.}} 
databases and then analysed them using IRAF 4\footnote{\url{http://iraf.noao.edu/.}}. 
In the case of IGR J17255-4509, SWIFT J1839.1-5717, and SWIFT J1937.5-4021 we stacked together the different spectra to increase the signal-to-noise ratio. As a result, we found that except for IGR J04085-6546, IGR J08297-4250, and SWIFT J1839.1-5717, all sources show optical features which are typical of AGN. In order to provide a classification, for these objects we examined the diagnostic ratio 
([OIII]$\lambda$5007/H$_{\beta}$ and [NII]$\lambda6583$/H$_{\alpha}$) according to \citet{1993ApJ...417...63H} and \citet{2003MNRAS.346.1055K}. The results of our findings are reported in Table \ref{tab3} where we list, for each source, the instrument used for the optical observation, the line ratios  used to determine the class, the measured redshift and AGN class. 
In this work, we report for the first time the optical class for 9 objects and the redshift value for 3 sources, namely IGR J11275-5319, IGR J13045-5630, and IGR J16560-4958 (see Table \ref{tab3}); for the remaining 6 galaxies, classified in this work, the redshifts we estimated are compatible with values in the literature. Of the nine sources with a clear optical class, 4 are LINERs and 5 are type 2 AGN. Interestingly, all  4 LINERs fall below the AGN threshold line of 0.5 in the WISE colour-colour diagram (Fig.\ref{fig3}). 
Although a contribution from absorption cannot be excluded,
as already mentioned before, this could indicate that the unusual WISE colours are due to radiatively inefficient accretion typically associated with low luminosity objects like LINERs.
As a further proof, we note that also NGC 4102, the only other LINER object in the present sample, has WISE colours (WI-W2=0.37 and W2-W3=4.1) that locate it in the same region of the diagram. On the other hand, this suggests that also IGR J16246-4556 and IGR J16459-2325, located below the AGN threshold in the WISE diagram of Fig. \ref{fig3}, and for which no optical spectrum is available, could be LINERs. 
Most of these LINERs are of type 2 with the possible exception of IGR J17255-4509, where a broad HeI line has been detected in the near infrared spectrum \citep{2022ApJS..261....8R} and indeed, as expected, most are absorbed in X-rays except for IGR J17255-4509.

As we mentioned before, 3 objects deserve more in-depth discussion as their optical spectra are somehow different to that expected from an emission line AGN.
The first source is IGR J04085-6546, associated with LEDA 310383:
although the diagnostics all point to an AGN, this source has been associated with  a transient event (OGLE 2013-SN-90) occurring on October 6, 2013 \citep{2013ATel.5488....1W}, originating from a galaxy and lasting several months.  The follow-up PESSTO observation confirmed the AGN class of the event \citep{2013ATel.5537....1S} despite the source photometric class pointed to a SN of type I \citep{2014AcA....64..197W}. The XRT observation used for this work was performed on Jan 3, 2013, after the peak of the transient event, while the INTEGRAL measurements are the average of multi-years observations.
Our analysis of the PESSTO spectrum, shows the presence of broad H$_{\alpha}$ and H$_{\beta}$ emissions and possibly a narrow 
[OIII]$\lambda$5007 emission line, at a redshift z=0.116$\pm$0.001.
However, since the spectrum was acquired about one month after the announcement of the OGLE transient, we cannot exclude a contribution from it, especially in the H$_{\beta}$ region. 
Nevertheless, we suggest a Seyfert 1 AGN classification for this source.

Another interesting source is IGR J08297-4250. In this case the low S/N X-Shooter spectrum does not allow us to infer the nature of this source. However, the absence of emission and absorption lines, especially at z=0, suggests this source is not a Galactic objects. 
The fact that the radio morphology of the source is core plus jet like (see section 4.1 and Fig. \ref{r1}, right) 
seems to point to a BL Lac class; the absence of emission lines in the optical spectra 
further validates this hypothesis and we therefore suggest that IGR J08297-4250 is a BL Lac with unknown redshift. Although it is difficult to discriminate blazars from  other AGN types  simply on the basis of two mid-IR colours, this becomes possible with the use of the three-dimensional WISE colour space, where a third colour, W3-W4, is also used (e.g Fig 6 in \citealt{2019ApJS..242....4D}); this extra information also allows to discriminate between blazar types, i.e. BL Lacs from Flat Spectrum Radio Quasars. In the case of IGR J08297-4250,  W3-W4 is 2.56, which locates the source in  the blazar region and  well inside the BL Lac locus.

A more controversial case is that of SWIFT J1839.1-5717: also in this source  no emission lines are detected in the optical spectrum suggesting a similar classification as for IGR J08297-4250. However we notice that \citet{2022ApJS..261....6K}, using the same optical spectrum, classify this source as a galactic object, i.e. a star, given the detection of a CaII absorption line. The object is located well below the Galactic plane, has WISE colours typical of an AGN and not of a stellar object and its optical spectrum
is characterised by a smooth continuum with no apparent absorption lines at redshift z=0. 
This gives us confidence to rule out the stellar nature for SWIFT J1839.1-5717 and given the above results, also in this case we suggest  the source classification as an AGN, and more specifically as a BL Lac.

\begin{figure}
\includegraphics[width=\textwidth,width=8cm]{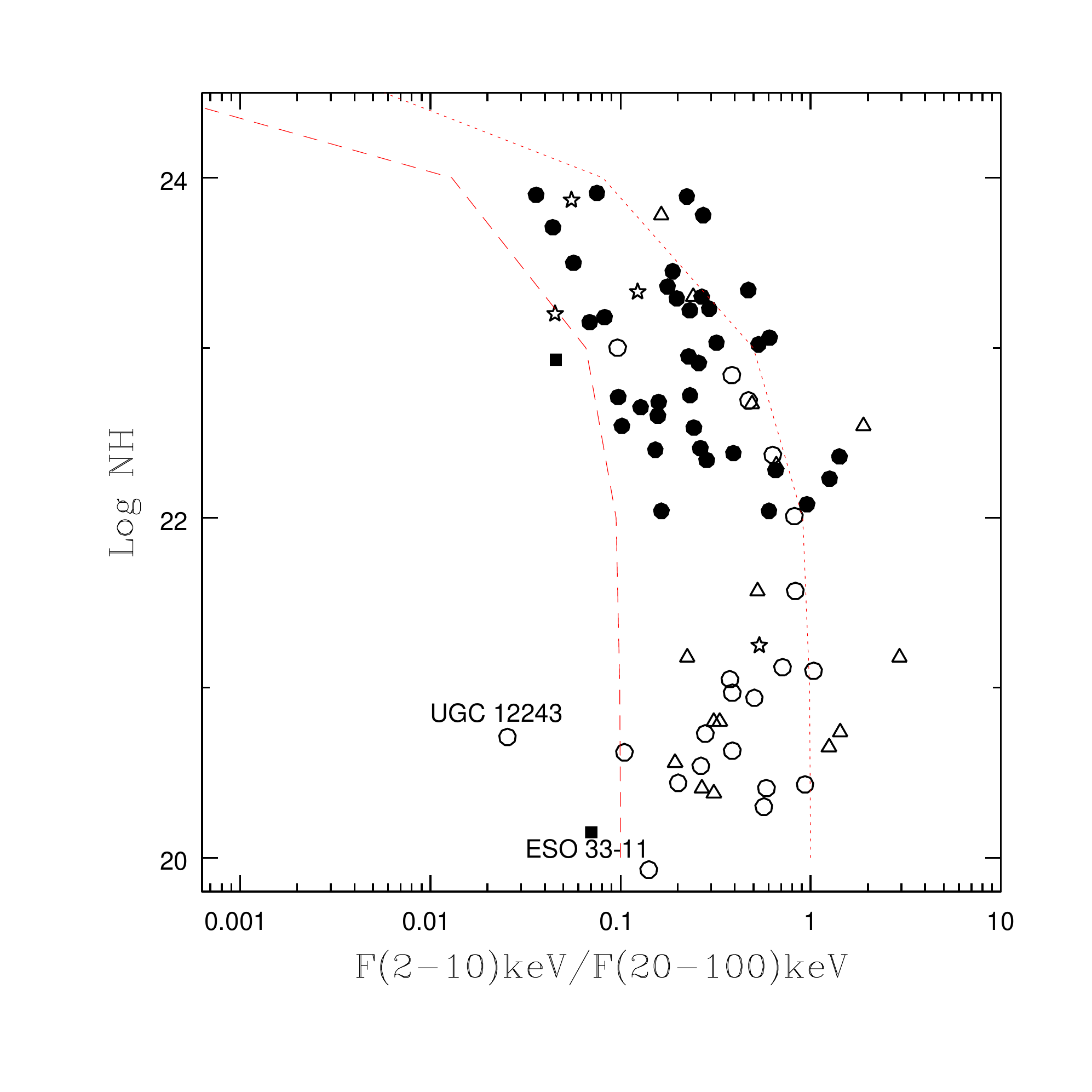}
\caption{Log NH versus F$_{2-10~keV}$/F$_{20-100~keV}$ flux ratio for our sample sources. Open circles are Seyfert 1-1.5 sources while filled circles are Seyfert 2 sources. Triangles are blazars, filled squares are XBONG and stars are LINERs.
Lines correspond as in \citet{2007ApJ...668...81M} to the expected values for an absorbed power law with photon index 1.5 (dotted line) and 1.9 (dashed line).} 
\label{fig4}
\end{figure}

\section{Discussion and Conclusions}
In this work we have added overall 83 objects to the previous INTEGRAL list of AGN: more than 40\% of them are in the zone of avoidance, i.e. behind our Galaxy  where INTEGRAL/IBIS  exposure is much deeper than Swift/BAT. This peculiarity makes the extragalactic nature of these objects more difficult to establish, indeed most of the 34 objects not yet classified in the literature  and listed as AGN in Table \ref{tab1}, are on the Galactic plane ($\sim$68\%).
For these 34 still unclassified sources we have used 3 diagnostics based on far-infrared data, radio detection and redshift measurement, position in the sky or extension in the optical images, in order to confirm their AGN nature.  
At least 2 diagnostics are verified for all the 34 candidates and therefore we can consider them AGN although of unknown class.
Indeed we were also able to classify 12 AGN among this set of candidates on the basis of public available optical spectra, finding that 5 of them are Seyfert 2, 4 are LINERs; one possible a Seyfert 1 galaxy (IGR J04085-6546) and two sources are possible BL Lac blazars (IGR J08297-4250 and SWIFT J1839.1-5717). Due to their still uncertain optical classes, we consider these last three objects as AGN candidates.  
Interestingly, the hard X-ray selection samples all types of AGN including the low luminosity ones. 
Indeed our sample lists 6 BL Lac objects (plus 2 possible), 6 Flat Spectrum Radio QSO, 1 Narrow Line Seyfert 1 galaxy, 5 LINERs, 2 XBONG or X-ray Bright Optically Normal Galaxies, the rest are Seyfert 1-1.5 (18) and Seyfert 1.8-2 (21). 

As previously observed by \citet{2012MNRAS.426.1750M}, the optical classification is generally reflected in the X-ray  absorption properties of the sources, i.e.  type 1 objects and blazars are generally  unabsorbed while type 2 sources are mildly/heavily absorbed.
A useful tool to identify outliers to this trend is the diagnostic diagram shown in Fig. \ref{fig4} as presented in \citet{2007ApJ...668...81M} where the source column density is plotted versus the soft to hard X-ray flux ratio (see \citealt{2007ApJ...668...81M} for a full description of  this diagnostic). In this plot a clear trend of a decreasing softness ratio as the absorption increases is visible, as expected if the 2 – 10 keV flux is progressively depressed as the absorption becomes stronger. 
Outliers could be absorbed type 1 AGN with column density above 10$^{22}$ cm$^{-2}$, strongly variable sources (Seyfert/blazar) where the soft and hard X-ray fluxes are not simultaneously measured and Compton thick objects where the column density is not properly estimated.
Using the entire INTEGRAL AGN sample, we find that these outliers are $\sim$10\%; some objects belonging to each class of these outliers have already been pinpointed and extensively discussed in previous works (see for example \citealt{2007ApJ...668...81M,2009MNRAS.394L.121M,2012MNRAS.426.1750M}).

Most of the sources in the present sample follow the expected trend, except for a few absorbed  type 1 AGN (i.e. open circles (Seyfert 1-1.5) and triangles (blazars) with column densities above 10$^{22}$ cm$^{-2}$) which fall in the region populated by absorbed objects.  
Their peculiarity may be due either to an incomplete account of the Galactic absorption especially in those AGN located  behind the Galactic plane (like for the two blazars 4C 50.11 and 4C 39.35) or to a poor quality of the X-ray spectrum (like for Fairall 1203). 
In  NGC 2617 and  MKN 817 the unexpected absorption is likely due to their spectral complexity: NGC 2617 is a changing look AGN with a complex X-ray absorption structure/broad line region \citep{2017A&A...597A..66G,2021MNRAS.503.3886Y}; MKN 817 is also quite peculiar, since its absorption is likely ionised and variable and probably interferes with the source  disk–corona  coupling \citep{2021ApJ...911L..12M}. 
A further anomalous case is that of AX J2254.7+1146(2)/UGC 12243, which as already discussed, is a type 1 AGN with no absorption but with  a very low soft to hard X-ray flux ratio. This could be due to strong variability in the high energy emission so that combining soft and hard X-ray data not taken contemporaneously, provides a false value of  the flux ratio. However, a comparison 
between XRT and XMM observations taken a few years apart, does not indicate a change in the source flux within the errors.
For this reason, the source may have a  highly absorbed soft X-ray spectrum which, if not properly corrected for the true column density value, provides again a wrong estimate of the flux ratio. 
In support of this last hypothesis we note that \citet{2022ApJS..261....4O} classify UGC 12243 as a type 2 AGN contrary to previous claim that it is a Seyfert 1 \citep{2010A&A...518A..10V}.
If this is true, then UGC 12243 could be another local Compton thick AGN, but
in order to understand the  true nature  of this source,  good quality X-ray observations are clearly needed.
As expected and visible in Fig. \ref{fig4}, none of the Seyfert 2 are unabsorbed, although the column density in several sources is often quite mild for a type 2 AGN. 

Interestingly we do not find any Compton thick AGN although we would have expected a few.
Following  \citet{2009MNRAS.399..944M}, 7\% of objects below redshift 0.335 are expected to be Compton thick (i.e. around 5  sources in  our sample);  more accurately \citet{2021ApJ...922..252T} estimate the fraction of Compton thick objects with redshift below 0.05  to be 7.7\%  or 2-3  objects in our case. 
This further confirms that also  at high energies absorption can prevent the detection of some heavily obscured sources as first reported in  \citet{2009MNRAS.399..944M}.
However, the diagnostic plot reported in Fig. \ref{fig4} shows a couple of objects which have a too low softness ratio values for the observed column densities, suggesting for them a Compton-thick nature; these are UGC 12243 (already discussed above) and the XBONG source IGR J05048-7340/ESO 33-11. In this last case  the optical classification as a normal galaxy could be due to the nuclear emission being completely hidden by the absorption; in addition, it must always be kept in mind that the poor quality of the X-ray spectral data can cause a wrong estimate of the X-ray column density. This is another interesting source that deserves more in depth studies.

The only Narrow Line Seyfert 1 reported in the sample is RX J1317.0+3735 \citep{2017ApJS..229...39R}, so far a poorly studied AGN; its black hole mass is log (M$_{\rm BH}$)=6.88, while its Eddington ratio based on the 5100 {\AA} monochromatic luminosity is log Edd=-0.53 \citep{2019ApJS..243...21L}, confirming that this class of AGN tends to have small black hole masses and high Eddington ratios \citep{2017ApJS..229...39R}.

Five objects in the present sample are classified as LINERs plus two, IGR J16246-4556 and IGR J16459-2325, likely to be LINERs candidates. Most of them display narrow lines and can therefore be classified as type 2 AGN while only one, IGR J17255-4509, likely belongs to the type 1 class. 
Their hard  X-ray luminosities range from 2 $\times$ 10$^{41}$ erg s$^{-1}$ to a few 10$^{43}$ erg s$^{-1}$. However, a couple of them, despite occupying the lower part of the  luminosity versus redshift diagram (see Fig. \ref{fig1}), cannot be considered  low luminosity AGN due to their high luminosity.
In addition, the Eddington ratios, available for only  a few objects, fall in the range 0.001-0.009  
\citep{2010A&A...521A..55C} indicating  again a behaviour much more similar to Seyfert galaxies. 
We conclude that the hard X-ray emitting LINERs are definitely active galaxies
but  whether their nucleus is similar or different from  more canonical AGN is still unclear. 
Finally, we note that  the WISE colour-colour diagram could be used to discriminate between LINERs and other AGN like Seyferts and blazars, as amply discussed in section 4.1.

Of the reported blazars, all are MeV emitters being detected by Fermi/LAT, with the only exceptions of IGR J01036-6439, IGR J04426-0018(2) and SWIFT J0600.7+0008.
Four blazars (SWIFT J0710.5+5908, RX J1136.5+6737, IGR J20596+4303 and 1ES 2344+514) are also TeV sources and all of these  are HBL or high frequency peaked BL Lac. 
None of the two  blazar candidates is a Fermi source.

In this work we have added to the INTEGRAL AGN catalogue 83 new sources of which more than 40\% are located behind the Galactic plane; this highlights the importance of INTEGRAL/IBIS in discovering new hard X-ray emitting AGN, mainly in this zone of avoidance where it is much more difficult to discriminate between galactic and extragalactic objects.
Their Galactic plane location makes the follow-up harder especially in the UV and optical bands, thus, 
the importance of also using  multiple diagnostic tests to properly classify each object.
In this paper we have shown that WISE colours, radio emission and morphological evidences can be used as powerful diagnostic tests to identify AGN candidates among high energy emitters.

\section*{Acknowledgements}
The authors acknowledge financial support from ASI under contract n.  2019-35-HH.0.

This research has made use of the NASA/IPAC Extragalactic Database (NED) and NASA/ IPAC Infrared Science Archive, which are operated by 
the Jet Propulsion Laboratory, California Institute of Technology, 
under contract with the National Aeronautics and Space Administration.
This work made use of observations collected at the European Southern Observatory under ESO programs 0103.A-0521(A), 0102.D-0918(A), 0102.A-0433(A), and 0101.A-0765(A).
The authors thank the anonymous referee for the useful comments.

\clearpage

\small
\begin{center}
\begin{table*}
\caption{AGN candidates} 
\begin{tabular}{|l c c|l l l|l |} 
                         &           \multicolumn{2}{c} {WISE colours} &  \multicolumn{3}{c} {RADIO FLUX} & Extra Data\\
\hline
Source	                 & W1 - W2  & W2 - W3   & SUMSS           & NVSS     & Other              & AGN marks     \\
                         &          &           & (0.843GHz)     & (1.4GHz)  &                    &               \\
 \hline\hline    
IGR J03574-6602          & 0.32     & 3.80      & 16.8$\pm$1.7   &  --       & --                 & z           \\
IGR J04059+5416          & 0.82     & 2.82      &        --      & --        & 1.92$\pm$0.30$^{a}$ & --   \\
IGR J04085-6546          & 0.72	    & 2.20      &   --           & --        & <33.0 $^{c}$       & z   \\
IGR J06075-6148          & 0.37     & 3.45      &  52.3$\pm$3.6  & --        &  --                & z  \\
IGR J06503-7742          & 0.25     & 2.29      &    --          &  --       & 5.10$\pm$0.51$^{c}$ & z   \\
IGR J08297-4250          & 0.54     & 3.09      &    --          &  --       & 4.81$\pm$0.48$^{c}$ & Ext  \\   
SWIFT J0835.5-0902       & 0.86     & 2.87      &    --          &  --       & 2.88$\pm$0.29$^{c}$ & z  \\  
SWIFT J0958.2-5732       & 1.14     & 2.28      & 13.2$\pm$1.9   &  --       &  --                 & z   \\ 
IGR J11275-5319          & 0.97     & 3.25      & 14.0$\pm$1.1   &  --       &  --                 & z \\
IGR J11299-6557          & 1.18     & 3.04      & --             &  --       & 5.98$\pm$0.60$^{c}$ & -- \\
IGR J13045-5630          & 0.98     & 3.26      & 16.8$\pm$1.4   &  --       & --                  & -- \\
PKS 1413-36              & 1.03     & 2.70      & --             & 1098.2$\pm$34.3 & --            & z \\
IGR J14417-5533          & 0.97     & 2.51      & --             & --        & <0.84$^{c}$         & Ext \\
IGR J14557-5448          & 0.81     & 3.13      & 33.1$\pm$1.4   & --        & --                  & --   \\
IGR J16005-4645          & 1.22     & 2.64      & 7843.0$\pm$235.3 & --      & --                  & -- \\
IGR J16181-5407          & 1.18     & 2.57      &   --           &  --       & 10.5$\pm$1.00$^{c}$  & z \\
IGR J16246-4556          & 0.18     & 1.78      &   --           &  --       & 9.51$\pm$0.95$^{c}$ &  Ext  \\
IGR J16413-4046          & 0.95     & 2.86      &   --           & --        & 2.23$\pm$0.22$^{c}$ & --   \\
IGR J16459-2325          & 0.28     & 2.18      &   --           & --        & <0.99$^{c}$         & z \\
IGR J16560-4958          & 0.67     & 2.19      &   --           & --        & 11.4$\pm$1.14$^{c}$ & Ext  \\
IGR J17157-5449          & 0.86     & 2.87      &   --           &  --       & 7.09$\pm$0.71$^{c}$ &  -- \\
IGR J17255-4509          & 0.36     & 3.00      & 10.9$\pm$1.2   &   --      &   --                & z  \\
IGR J18134-1636          & 0.74     & 3.49      &  --            &  --       & 1.98$\pm$0.40$^{a}$  &  --  \\
IGR J18141-1823          & 0.73     & 2.99      &  --            &  --       & 10.3$\pm$6.80$^{b}$  & -- \\
AXJ 1830.6-1002          & 1.02     & 1.87      &  --            &  --       & 2.54$\pm$0.40$^{a}$  & --  \\
SWIFT J1839.1-5717       & 1.84	    & 3.47      &  --            &  --       &<0.57$^{c}$          &l$\le$-20  \\
IGR J18486-0047          & 1.72     & 2.68      &  --            & 20.6$\pm$1.7  &  --             & --  \\
IGR J18497-0248          & 0.66     & 2.51      &  --            & 15.9$\pm$1.6  &  --             & --  \\
IGR J19294+1328          & 0.89     & 2.57      &  --            & --            & 3.42$\pm$0.30$^{a}$ & --   \\
SWIFT J1937.5-4021-1     & 0.77     & 2.96      &  --            & --            & 4.87$\pm$0.49$^{c}$ &  z  \\
SWIFT J1937.5-4021-2     & 0.81     & 2.40      &  --            & --            & <1.05 $^{c}$        & l$\le$-20  \\
IGR J19504+3319          & 0.85     & 2.18      &  --            & --            & 3.972$\pm$0.2$^{a}$ & -- \\
IGR J19577+3339          & 0.8*     & --        & --             & 295.2$\pm$8.9 & --               & -- \\
SWIFT J2055.0+3559       & 1.38     & 2.98      &  --            &  9.6$\pm$0.5  &  --              & z \\
\hline   
\multicolumn{7}{l}{a= VLASS;  b = MAGPIS;  c= RACS}\\
\label{tab2}
\end{tabular}
\end{table*}
\end{center}

\small
\begin{center}
\begin{table*}
\caption{Confirmed AGN: optical spectra}
\begin{tabular}{lccclcc}
\hline
Source              & Instrument     &  Log(OIII$\lambda5007$/H$_{\beta}$)  & Log(NII$\lambda6583$/H$_{\alpha}$)  & H$_{\alpha}$/H$_{\beta}$   &$z$  & class     \\
\hline
IGR J03574-6602       & X-shooter      &  0.42                  & -0.120         & 5.73       & 0.0190   & LINER \\
IGR J04085-6546       & PESSTO         &   -                    &   -            & -          &          &  Sy 1?    \\
IGR J06075-6148       & 6dF            & -0.014                 & -0.350         & 2.88       & 0.0040   & LINER \\
IGR J06503-7742       & 6df            & 0.057                  &  0.511         & 12.64      & 0.0373   & LINER \\
IGR J08297-4250       & X-shooter      &  -                     &   -            &   -        &          & Blazar/BL Lac?\\
IGR J11275-5319       & X-shooter      & 1.01                   & -0.062         & 3.58       & 0.0503   & Sy 2 \\
IGR J13045-5630$^{a}$ & X-shooter      & 0.91                   & -0.140         & 5.00       & 0.0511   & Sy 2 \\
PKS 1413-36           & X-shooter      & 1.00                   & -0.313         & 3.75       & 0.0753   & Sy 2 \\
IGRJ16560-4958$^{a}$  & X-shooter      & 1.20                   & 0.460          & $>$10.77   & 0.0586   & Sy 2 \\   
IGR J17255-4509$^{a}$ & X-shooter      & 0.80                   & 0.002          & 10.40      & 0.0197   & LINER \\
SWIFT J1839.1-5717    & X-shooter      &  -                     &  -             & -          &     -    & Blazar/BL Lac?    \\
SWIFT J1937.5-4021    & X-shooter      & 1.36                   & 0.400          & 8.59       & 0.0740    & Sy 2 \\              
\hline   
\multicolumn{7}{l}{a = for this source we report the redshift estimate for the first time}\\
\label{tab3}
\end{tabular}
\end{table*}
\end{center}



\bibliographystyle{aa}
\bibliography{agn-up}


\appendix
\onecolumn
\section{New INTEGRAL/IBIS AGN}

\begin{landscape}
  \begin{tablenotes}  
  \small
     \item[]   {\bf Notes}: Sources in bold are (also) in cat 1000; values of Log N$_{H}$ in bold are for sources where only Galactic absorption is measured, for the rest of the sample sources the Galactic absorption has been always considered in addition to the intrinsic one listed; $^{(\dagger)}$ 2-10 keV flux in units of 10$^{-12}$ erg cm$^{-2}$ s$^{-1}$;  $^{(\ddagger)}$ 20-100 keV flux in units of 10$^{-11}$ erg cm$^{-2}$ s$^{-1}$;    
      {\bf(1)} Two sources are present within the IBIS error circle: a Seyfert 1 galaxy, whose values are reported in the table, and an eclipsing X-ray binary, a transitional millisecond pulsar (SXPS J042749.2-670434). Both sources are detected above 3 keV, therefore the hard X-ray emission could be due to the contribution of the two;
      {\bf(2)} Two sources present within the IBIS error circle, both AGN, both reported in the table. The hard X-ray emission could be the result 
      of the contribution of two objects and therefore the value reported in Krivonos has been divided by 2;
     {\bf(3)} Two sources lie within the IBIS error circle: the galaxy  2MASX J08353333-0905302' with the photometric redshift reported in the table, and a rotationally variable Star (ASAS J083531-0904.1). Both objects emit above 5 keV, therefore they could both contribute to the hard X-ray emission.
     {\bf $^{(\star)}$} For this sources only the XMM Slew Survey flux is available;
     {\bf (4$^{\star})$} Two sources are listed in the XMM Slew Survey: one is LEDA140000 (also ROSAT Bright) reported in the table, while the other, XMMSL2 J124313.8+780828 with a 2-10 keV flux of 5.9 $\times$ 10$^{-12}$ erg cm$^{-2}$ s$^{-1}$, is unidentified. Probably, they both contribute to the high-energy emission;
    {\bf(5)} Two sources lie within the IBIS error circle: one is the Seyfert 2 reported in the table while the second is a CV. 
    They both contribute to the high-energy flux;
   {\bf $^{\diamond}$} Mean value; {\bf $^{\clubsuit}$} Photometric redshift; {\bf $^{\ast}$} Redshift and optical class from \citet{2022ApJS..261....4O}; 
                          \item[]         {\bf References}:   
                                     1): this work; 
                                     2): \citet{2021ApJS..257...61Y};
                                     3): \citet{2017ApJS..233...17R}; 
                                     4): \citet{2015MNRAS.449..597T}; 
                                    7): \citet{2020MNRAS.497..229A};
                                    8) \citet{2008A&A...480..611S}; 
                                    9) \citet{2021A&A...650A..57Z}; 
                                    10) \citet{2017A&A...597A..66G};
                                    11) \citet{2017MNRAS.470.1107L};
                                    12) \citet{2019ApJ...887...32C}; 
                                    13) \citet{2010ATel.2557....1R};
                                    14) \citet{2021ApJ...922..151K};
                                    15) \citet{2020ApJ...889...53T}; 
                                    16) \citet{2011ATel.3178....1L};
                                    17) \citet{2012ApJ...754..145T}
                                    18) \citet{2009MNRAS.395L...1B};
                                    19) \citet{2017MNRAS.465.1563R};
                                    20) \citet{2021ApJ...914...48T}
                                    21) \citet{2020AstL...45..836K}
\end{tablenotes}

\small
\begin{longtable}{l l c c  c c c c c c c }
\caption{INTEGRAL/IBIS  AGN}\\
\hline
 Name	                     & altern. name         &  RA                 & dec                   &  z          & class          & Log N$_{H}$                       & $\Gamma_{2-10~keV}$  &    F$_{S}^{\dagger}$&  F$_{H}^{\ddagger}$  & 	Ref.	\\
 \hline
 \endfirsthead
\multicolumn{11}{c}%
 {\tablename\ \thetable\ -- \textit{Continued from previous page}} \\
  \hline
 Name	                     & altern. name  &  RA                 & dec                   &  z          & class          & Log N$_{H}$                       & $\Gamma_{2-10~keV}$  &    F$_{S}^{\dagger}$&  F$_{H}^{\ddagger}$  & 	Ref.	\\ 
 \hline
\endhead
\hline \multicolumn{11}{c}{\textit{Continued on next page}} \\
\endfoot
\hline
\endlastfoot
SWIFT J0001.6-7701            & Fairall 1203             & 00 01 46.08   & $-$76 57 14.3   &  0.058  & Sy 1          &  22.69$^{+0.10}_{-0.10}$  & 1.8 (fixed)               &   3.39       &  0.72    &      1 (XRT) \\
NGC 235A                      &	  -		                 & 00 42 52.81   & $-$23 32 27.7   &  0.022  & Sy 2          &  23.71$^{+0.06}_{-0.06}$  & 1.68$^{+0.11}_{-0.10}$    &   2.10       &  4.76    &      2 \\
IGR J00569+6359               & 3PBC J0056.9+6401		 & 00 57 12.70   & $+$63 59 53.2   &  0.291  & Sy 1.2        &  22.37$^{+0.07}_{-0.08}$  & 1.79$^{+0.21}_{-0.25}$    &   2.40       &  0.38    &      3 \\
IGR J01036-6439               & PKS 0101-649             & 01 03 33.74   & $-$64 39 07.7   &  0.163  & Blazar/QSO    & {\bf 20.41}               & 1.66$^{+0.08}_{-0.06}$    &   3.00       &  1.12    &      3 \\ 
IGR J01242+3348               & NGC 513                  & 01 24 26.80   & $+$33 47 58.2   &  0.020  & Sy 2          &  22.72$^{+0.04}_{-0.05}$  & 1.47$^{+0.07}_{-0.07}$    &   3.89       &  1.68    &      1 (XMM + NuSTAR) \\
SWIFT J0238.2-5213            & ESO 198-24               & 02 38 19.72   & $-$52 11 32.3   &  0.045  & Sy 1          & {\bf 20.43}               & 1.72$^{+0.14}_{-0.11}$    &  12.9        &  1.38    &      3 \\ 
SWIFT J0250.2+4650            & LEDA 2287192             & 02 50 27.18   & $+$46 47 29.4   &  0.0210  & Sy 2          & 22.38$^{+0.10}_{-0.09}$   & 1.36$^{+0.28}_{-0.26}$    &   6.40       &  1.63    &      1 (XRT) \\  
IGR J03574-6602               & NGC 1503                 & 03 56 33.14   & $-$66 02 27.6   &  0.019  & AGN           & $\sim$23.26               & 1.8 (fixed)               &   0.34       &  0.75    &      1 (XRT) \\ 
SWIFT J0359.7+5058            & 4C 50.11                 & 03 59 29.75   & $+$50 57 50.2   &  1.520  & Blazar/QSO    & 22.67$^{+0.12}_{-0.10}$   & 1.47$^{+0.20}_{-0.07}$    &   5.40       &  1.10    &      3 \\
IGR J04059+5416               & 2MASX J04055765+5418446  & 04 05 57.66   & $+$54 18 44.7   &   --    & AGN           & 22.54$^{+0.23}_{-0.34}$   & 1.50$^{+0.90}_{-0.80}$    &   1.10       &  1.08    &      4 \\    
IGR J04085-6546               & LEDA 310383              & 04 08 38.82   & $-$65 45 59.1   &  0.125  & AGN           & {\bf 20.56}               & 1.63$^{+0.31}_{-0.30}$    &   1.31       &  0.67    &      1 (XRT) \\
SWIFT J0414.8-0754            & LEDA 14727               & 04 14 52.66   & $-$07 55 39.7   &  0.038  & Sy 1          & {\bf 20.73}               & 1.97$^{+0.92}_{-0.27}$    &   5.00       &  1.79    &      3 \\
NGC 1566                      &   -                      & 04 20 00.40   & $-$54 56 16.6   &  0.005  & Sy 1          & {\bf 19.93}               & 1.73$^{+0.07}_{-0.06}$    &   3.10       &  2.20    &      3 \\
SWIFT J0427.0+0734            & 2MASS J04270427+0716316  & 04 27 04.28   & $+$07 16 31.7   &  0.096  & Sy 1          & {\bf 20.94}               & 1.61$^{+0.17}_{-0.17}$    &   4.20       &  0.83    &      1 (XRT)\\
IGR J04288-6702$^{(1)}$       & LEDA 299570              & 04 29 47.48   & $-$67 03 18.9   &  0.065  & Sy 1.5        & {\bf 20.54}               & 1.64$^{+0.11}_{-0.11}$    &   1.78       &   0.68   &      1 (XRT) \\
IGR J04426-0018$^{(2)}$       & 1RXS J044229.8-001823    & 04 42 30.18   & $-$00 18 29.1   &  0.449  & Blazar/BL Lac & {\bf 20.80}               & 1.83$^{+0.07}_{-0.07}$    &   1.39       &   0.42   &      1 (XRT)\\
IGR J04426-0018$^{(2)}$       & PKS 0440-003             & 04 42 38.59   & $-$00 17 42.6   &  0.845  & Blazar/QSO    & {\bf 20.80}               & 1.92$^{+0.09}_{-0.09}$    &   1.32       &   0.42   &      1 (XRT)\\
IGR J05048-7340               & ESO 33-11                & 05 05 06.27   & $-$73 39 04.6   & 0.014   & XBONG         & {\bf 20.15}                & 1.8$^{+0.3}_{-0.3}$           &  0.33      &   0.47   &      1 (XRT)\\
IGR J05162-1034               & MCG-02-14-009            & 05 16 21.22   & $-$10 33 41.4   &  0.029  & Sy 1          & {\bf 20.97}               & 2.23$^{+0.05}_{-0.05}$    &   4.20       &  1.08    &      3 \\
{\bf IGR J05511-1218}         & LEDA 148076              & 05 51 13.12   & $-$12 14 42.1   & 0.035   & XBONG         & 22.93$^{+0.32}_{-0.24}$   & 1.8 (fixed)               &   0.8        & $<$1.75  &      1 (XRT), 5, 6 \\
SWIFT J0600.7+0008            & IRAS05581+0006           & 06 00 40.11   & $+$00 06 18.4   &  0.115  & Sy 1.9    & 23.30$^{+0.11}_{-0.11}$   & 1.8 (fixed)               &   1.99       &  0.83    &      1 XRT \\ 
IGR J06075-6148               & ESO 121-6                & 06 07 29.85   & $-$61 48 27.3   &  0.004  & AGN           & 23.33$^{+0.03}_{-0.03}$   & 1.89$^{+0.08}_{-0.08}$    &   0.95       &  0.73    &      7 \\ 
IGR J06380-7536               & LEDA 243576              & 06 37 43.17   & $-$75 38 46.3   & 0.089   & Sy 1.8        &  --                       &           --              &   --         &  0.78    &      no (2-10) data \\
IGR J06503-7742$^{\star} $    & LEDA 235040              & 06 49 54.37   & $-$77 42 18.6   & 0.037   & AGN           &  --                       &           --              &   0.64       &  0.94    &      8 \\ 
IGR J07072-1227               & 2MASX J07071126-1227560  & 07 07 11.43   & $-$12 28 00.2   & 0.071   & Sy 2          & 22.71$^{+0.56}_{-0.35}$   & 1.8 (fixed)               &   1.11       & 1.13     &      1 (XRT), 5 \\
SWIFT J0710.5+5908            & RX J0710.5+5908          & 07 10 30.07   & $+$59 08 20.3   & 0.125   & Blazar/BL Lac & {\bf 20.65}               & 1.80$^{+0.04}_{-0.04}$    &  28.5$^{\diamond}$ & 2.28 &    1 (XRT) \\ 
IGR J07328-4640               & PKS 0731-465             & 07 32 44.31   & $-$46 40 17.2   & --      & Blazar/QSO      & 21.18$^{+0.20}_{-0.28}$   & 1.8 (fixed)               &   1.99       &  0.89     &     1 (XRT)\\ 
RX J0818.9-2252               & LEDA 80921               & 08 18 57.72   & $-$22 52 36.2   & 0.035   & Sy 1          & {\bf 21.05}               & 1.79$^{+0.09}_{-0.09}$    &   3.46       &  0.93     &     1 (XRT)\\ 
SWIFT J0823.4-0457            & Fairall 272              & 08 23 01.10   & $-$04 56 05.4   & 0.022   & Sy 2          & 23.23$^{+0.05}_{-0.05}$   & 1.51$^{+0.11}_{-0.08}$    &   6.46       &  2.22     &     9 \\
IGR J08297-4250               & 2MASX J08294112-4251582  & 08 29 41.12   & $-$42 51 58.3   &  --     & AGN           & 23.78$^{+0.43}_{-0.53}$   & 1.50$^{+0.90}_{-0.80}$    &   1.10       &  0.67     &     4 \\
SWIFT J0835.5-0902$^{(3)}$    & 2MASX J08353333-0905302  & 08 35 33.34   & $-$09 05 30.2   & 0.112$^{\clubsuit}$ & AGN &  {\bf 20.62}            & 2.04$^{+0.13}_{-0.13}$    &   1.68       &  1.62     &     1 (XRT)\\ 
NGC 2617                      &   -                      & 08 35 38.00   & $-$04 05 17.9   & 0.014   & Sy 1/1.8      & 23.34$^{+0.07}_{-0.08}$   & 1.87$^{+0.01}_{-0.01}$    &  23.5        &  5.00     &     10  \\ 
PKS 0921-213                  &   -                      & 09 23 38.88   & $-$21 35 47.1   & 0.053   & Sy 1          & {\bf 20.63}               & 1.77$^{+0.08}_{-0.08}$    &   6.23       &  1.60     &     1 (XRT)\\ 
{\bf SWIFT J0924.2-3142}      & 2MASS J09235373-3141308 &  09 23 53.61   & $-$31 41 31.6   & 0.042   & Sy 1.8        & 23.91$^{+0.23}_{-0.27}$    & 1.8 (fixed)               &   1.7   &  2.26      & 11 \\
SWIFT J0958.2-5732            & WISEA J095834.97-572927.2 & 09 58 35.03  & $-$57 29 26.7   & --      & AGN           & 22.28$^{+0.07}_{-0.09}$   & 1.8 (fixed)               &   3.44       &  0.52     &     1 \\ 
ESO 317-41                    & SWIFTJ1031.5-4205        & 10 31 23.11   & $-$42 31 23.1   & 0.019$^{\ast}$ & Sey 2 & 23.90$^{+0.07}_{-008.}$   & 1.76$^{+0.31}_{-0.21}$    &   0.44       &  1.11     &     9 \\
IGR J10447-6027               & 2MASS J10445192-6025115  & 10 44 51.91   & $-$60 25 11.9   & 0.047   & Sy 2         & 23.22$^{+0.07}_{-0.07}$   & 1.50$^{+0.06}_{-0.06}$    &   2.20       &  0.95     &     12 \\
IGR J10595-5125               & ESO 215-14               & 10 59 19.03   & $-$51 26 32.3   & 0.019   & Sy 1          & {\bf 21.10}               & 1.65$^{+011.}_{-0.11}$    &   5.71       &  0.55     &     1 (XRT)\\ 
IGR J11275-5319               & 2MASX J11273392-5320270  & 11 27 33.90   & $-$53 20 25.9   & 0.049$^{\clubsuit}$ & AGN & 23.18$^{+0.26}_{-0.22}$ & 1.8 (fixed)               &   0.84       &  0.97     &     1 (XRT) \\
IGR J11299-6557               & 2MASS J11295643-6555218  & 11 29 56.90   & $-$65 55 18.9   &  --     & AGN          & $<$22.04                  & 2.09$^{+0.98}_{-0.81}$    &   3.20       &  0.53     &     1 (XRT)\\ 
RX J1136.5+6737               & SWIFT J1136.7+6738       & 11 36 30.10   & $+$67 37 04.0   & 0.134   & Blazar/BL Lac & 20.74$^{+0.09}_{-0.14}$   & 2.04$^{+0.05}_{-0.04}$    &  11.6        &  0.81     &     3 \\
NGC 4102                      &  -                       & 12 06 23.11   & $+$52 42 39.4   & 0.003   & LINER         & 23.87$^{+0.04}_{-0.03}$   & 1.74$^{+0.12}_{-0.08}$    &   1.44       &  2.53     &     9 \\   
IGR J12418+7805$^{4,\star} $  &  LEDA 140000             & 12 42 36.10   & $+$78 07 20.4   & 0.022   & Sy 1.9        &      --                   &    --                     &   1.00       &  0.94     &     8 \\
PKS 1252+11                   &   -                      & 12 54 38.25   & $+$11 41 05.9   & 0.872   & Blazar/QSO    & {\bf 20.38}               & 1.30$^{+0.48}_{-0.48}$    &   1.77       &  0.58     &     1 (XRT)\\ 
IGR J13045-5630               & WISE J130431.77-563058.5 & 13 04 31.50   & $-$56 30 54.8   & --      & AGN           & 23.29$^{+0.24}_{-0.28}$   & 1.8 (fixed)               &   1.82       &  0.91     &     13 \\
RX J1317.0+3735               &    -                     & 13 17 02.89   & $+$37 35 32.7   & 0.195   & NLS1          &     --                    &    --                     &              &  1.02     &     no (2-10) keV data \\
ESO 509-G038                  & -                        & 13 31 13.84   & $-$25 24 09.9   & 0.026   & Sy 1          & {\bf 20.44}               & 1.56$^{+0.09}_{-0.09}$    &   3.44       &  1.69     &     1 (XRT)\\ 
PKS 1413-36                   & -                        & 14 16 33.16   & $-$36 40 53.8   & 0.075   & AGN           & 22.53$^{+0.13}_{-0.11}$   & 1.8 (fixed)               &   1.77       &  0.74     &     1 (XRT)\\    
MRK 1383                      &  -                       & 14 29 06.57   & $+$01 17 06.1   & 0.087   & Sy 1          & {\bf 20.41}               & 1.87$^{+0.02}_{-0.02}$    &   6.50       &  1.11     &     1 (XMM+NuSTAR)\\  
MRK 817                       &   -                      & 14 36 22.08   & $+$58 47 39.4   & 0.031   & Sy 1.2        & 22.84$^{+0.05}_{-0.04}$   & 1.91$^{+0.04}_{-0.09}$    &   6.67       &  1.74     &     14 \\
MRK 477                       & -                        & 14 40 38.09   & $+$53 30 16.2   & 0.038   & Sy 2          & 23.30$^{+0.06}_{-0.07}$   & 1.58$^{+0.09}_{-0.07}$    &   3.38       &  1.27     &     9 \\
IGR J14417-5533               & WISE J144118.74-553335.1 & 14 41 18.94   & $-$55 33 33.5   &  --     & AGN           & {\bf 21.57}               & 1.55$^{+0.24}_{-0.24}$    &   7.38       &  0.89     &     1 (XRT) \\
IGR J14557-5448               & LEDA 415943              & 14 55 32.19   & $-$54 46 31.3   & --      & AGN           & 23.15$^{+0.17}_{-0.12}$   & 1.8 (fixed)               &   0.44       &  0.58     &     1 (XMM)\\
IGR J16005-4645               & PMN J1600-4649           & 16 00 20.40   & $-$46 48 41.3   & --      & AGN          & 22.08$^{+0.39}_{-0.51}$   & 1.8 (fixed)               &   4.40       &  0.46     &     1 (XRT)\\
3C 332.0                      & -                        & 16 17 42.54   & $+$32 22 34.4   & 0.151   & Sy 1          & 22.01$^{+0.17}_{-0.10}$   & 1.77$^{+0.04}_{-0.04}$    &   9.22       &  1.12     &     1 (XRT + NuSTAR)\\ 
IGR J16181-5407               & 2MASS J16180771-5406122  & 16 18 07.74   & $-$54 06 12.2   & 0.085   & AGN           & 22.95$^{+0.25}_{-0.65}$   & 0.8$^{+1.10}_{-1.10}$     &   1.27       &  0.57     &     15 \\
{\bf IGR J16246-4556}         & 2MASS J16243080-4555144  & 16 24 30.77   & $-$45 55 14.1   & --      & AGN           & 22.40$^{+0.21}_{-0.45}$   & 1.3$^{+0.70}_{-0.07}$     &   0.93       & $<$0.59   &     15 \\
IGR J16413-4046               & CXOU J164119.4–404737    & 16 41 19.5    & $-$40 47 37.7   & --      & AGN           & 23                        &    1.8 (fixed)            &     1.40     &   1.45    &     16     \\  
IGR J16459-2325               & ESO 518-2                & 16 45 54.95   & $-$23 27 05.5   & 0.020   & AGN           & $\sim$23.78               & 1.8 (fixed)               &   4.30       &  1.58     &     1 (XRT) \\  
IGR J16560-4958               & CXOU J165551.9–495732    & 16 55 51.95   & $-$49 57 32.4   & --      & AGN           & 22.36$^{+0.07}_{-0.11}$   & 2.3$^{+0.70}_{-0.40}$     &  17.00       &  1.20     &     17 \\
IGR J17157-5449               & 2MASS J17153752-5450062  & 17 15 37.53   & $-$54 50 05.1   & --      & AGN          & {\bf 21.12}               & 1.51$^{+0.15}_{-0.15}$    &   4.25       &  0.59     &     1 (XRT)\\
IGR J17255-4509               & 2MASX J17253053-4510279  & 17 25 30.55   & $-$45 10 27.3   & 0.019   & AGN           & 21.25$^{+0.24}_{-0.30}$   & 1.8 (fixed)               &   2.91       &  0.54     &     1 (XRT) \\
SWIFT J1745.4+2906            & 1RXSJ174538.1+290823     & 17 45 38.29   & $+$29 08 22.7   & 0.111   & Sy 1          & $<$20.30                  & 1.57$^{+0.03}_{-0.03}$    &   7.08       &  1.25     &     1 (XRT+ NuSTAR)\\
IGR J18134-1636               & -                        & 18 13 28.03   & $-$16 35 48.5   & --      & AGN          & 22.65$^{+0.39}_{-0.33}$   & 1.8 (fixed)               &   1.15       &  0.94     &     1 (XRT)\\ 
IGR J18141-1823               & 4PBC J1814.1-1822        & 18 14 14.78   & $-$18 23 09.6   &  --     & AGN          & 22.68$^{+0.21}_{-0.19}$   & 1.8 (fixed)               &   0.98       &  0.63     &     1 (XRT)\\ 
AX J1830.6-1002               &   -                      & 18 30 38.3    & $-$10 02 47.1   & --      & AGN          & 23.03$^{+0.12}_{-0.13}$   & 1.01$^{+0.57}_{-0.38}$    &   3.20       &  1.00     &     18\\
IGR J18381-0924               & CXOU J183818.5-092552    & 18 38 18.58   & $-$09 25 52.2   & 0.031   & Sy 1.9        & 22.34$^{+0.03}_{-0.03}$   & 1.19$^{+0.07}_{-0.07}$    &   2.30       &  0.81     &     19 \\
{\bf SWIFT J1839.1-5717}      & allWISE J183905.95-571505.1& 18 39 06.37 & $-$57 15 05.8   &  --     & AGN           & 22.31$^{+0.07}_{-0.07}$   & 1.57$^{+0.21}_{-0.21}$    &   8.40       &  1.27     &     11 \\
IGR J18486-0047               & CXOU J184825.4?004635    & 18 48 25.45   & $-$00 46 34.9   &  --     & AGN           & 23.45$^{+0.17}_{-0.28}$  & 1.8 (fixed)                &   1.63       &  0.85     &     1 (XRT)\\ 
IGR J18497-0248               &  NVSS J184946-024819     & 18 49 46.67   & $-$02 48 17.3   & --      & AGN           & $\sim$22.60               & 1.8 (fixed)               &   1.08       &  0.70     &     1 (XRT)\\
IGR J19294+1328               & CXOU J192930.1+132705    & 19 29 30.11   & $+$13 27 05.6   & --      & AGN          & 23.89$^{+0.25}_{-0.21}$   & 1.8 (fixed)               &   1.87       &  0.85     &     1(XRT), 14 \\ 
SWIFT J1937.5-4021$^{2}$      & LEDA 588288              & 19 37 13.47   & $-$40 16 14.8   & 0.075   & AGN           & 22.23$^{+0.09}_{-0.10}$   & 1.51$^{+0.25}_{-0.25}$    &   3.92       &  0.31     &     1(XRT) \\            
SWIFT J1937.5-4021$^{2}$      & 1RXS J193716.1-401026    & 19 37 14.70   & $-$40 10 14.9   &   --    & AGN           & 22.91$^{+0.46}_{-0.67}$   & 1.92$^{+0.21}_{-0.23}$    &   0.78       &  0.31     &     1(XRT) \\ 
IGR J19504+3319               & 2MASS J19501973+3314166  & 19 50 19.73   & $+$33 14 16.2   & --      & AGN          & 22.04$^{+0.14}_{-0.19}$   & 1.8$^{+0.3}_{-0.3}$       &   2.24       &  1.34     &     20 \\
IGR J19577+3339               & CXOU J195740.5+333828    & 19 57 40.59   & $+$33 38 28.2   & --      & AGN          & $<$23.5                   & -0.6$^{+4.5}_{-1.6}$      &   0.26       &  0.53     &     20 \\
IC 5063                       & -                        & 20 52 02.33   & $-$57 04 07.6   & 0.011   & Sy 2          & 23.36$^{+0.02}_{-0.02}$   & 1.77$^{+0.08}_{-0.07}$    &   8.70       &  4.92     &      9 \\
SWIFT J2055.0+3559            & 2MASS J20550835+3556278  & 20 55 08.23   & $+$35 56 27.4   & 0.115$^{\clubsuit}$ & AGN & 23.02$^{+0.16}_{-0.15}$ & 1.8 (fixed)               &   2.50       &  0.47     &      1 (XRT) \\
IGR J20569+4940               & 4C 49.35                 & 20 56 42.74   & $+$49 40 06.6   & 0.100$^{\clubsuit}$ & Blazar/BL Lac & 22.54$^{+0.05}_{-0.05}$ & 2.60$^{+0.03}_{-0.03}$  & 22.0 &  1.16     &      12 \\
IGR J20596+4303$^{(5)}$       & -                        & 21 00 01.00   & $+$43 02 11.0   & 0.066   & Sy 2          & 22.41$^{+0.11}_{-0.13}$   & 1.8                       &   1.49       &  0.57     &      21 \\
IGR J22018+5049               & NRAO 676                 & 22 01 43.54   & $+$50 48 56.4   & 1.899   & Blazar/QSO    & {\bf 21.57}               & 1.39$^{+0.07}_{-0.07}$    &   5.20       &  0.99     &      1 (XRT) \\ 
AX J2254.3+1146$^{(2)}$       & UGC 12237                & 22 54 19.67   & $+$11 46 56.8   & 0.0286  & Sy 2          & 23.06$^{+0.14}_{-0.13}$   & 1.8 (fixed)               &   3.10       &  0.51     &      1 (XRT)\\ 
AX J2254.7+1146$^{(2)}$       & UGC 12243                & 22 54 43.51   & $+$11 42 50.9   & 0.0278  & Sy 1          & {\bf 20.71}               & 1.75$^{+0.72}_{-0.79}$    &   0.13       &  0.51     &      1 (XRT)\\
1ES 2344+514                  & -                        & 23 47 04.84   & $+$51 42 17.9   & 0.044   & Blazar/BL Lac & {\bf 21.18}               & 1.96$^{+0.02}_{-0.02}$        &  17.0        &  0.58     &      1 (XRT) \\ 
\label{tab1}
\end{longtable}
\end{landscape}

\section{Swift/XRT fields of multiple counterparts}
In this Appendix, the Swift/XRT fields of peculiar sources where multiple counterparts have been found and discussed in Section 3, are shown. To visualise better and highlight the X-ray counterparts of the new INTEGRAL AGN, the images have been smoothed, therefore the presence of grains and/or features inside the XRT field of view are undoubtedly spurious.


\begin{figure*}
\includegraphics[width=\textwidth,width=8cm]{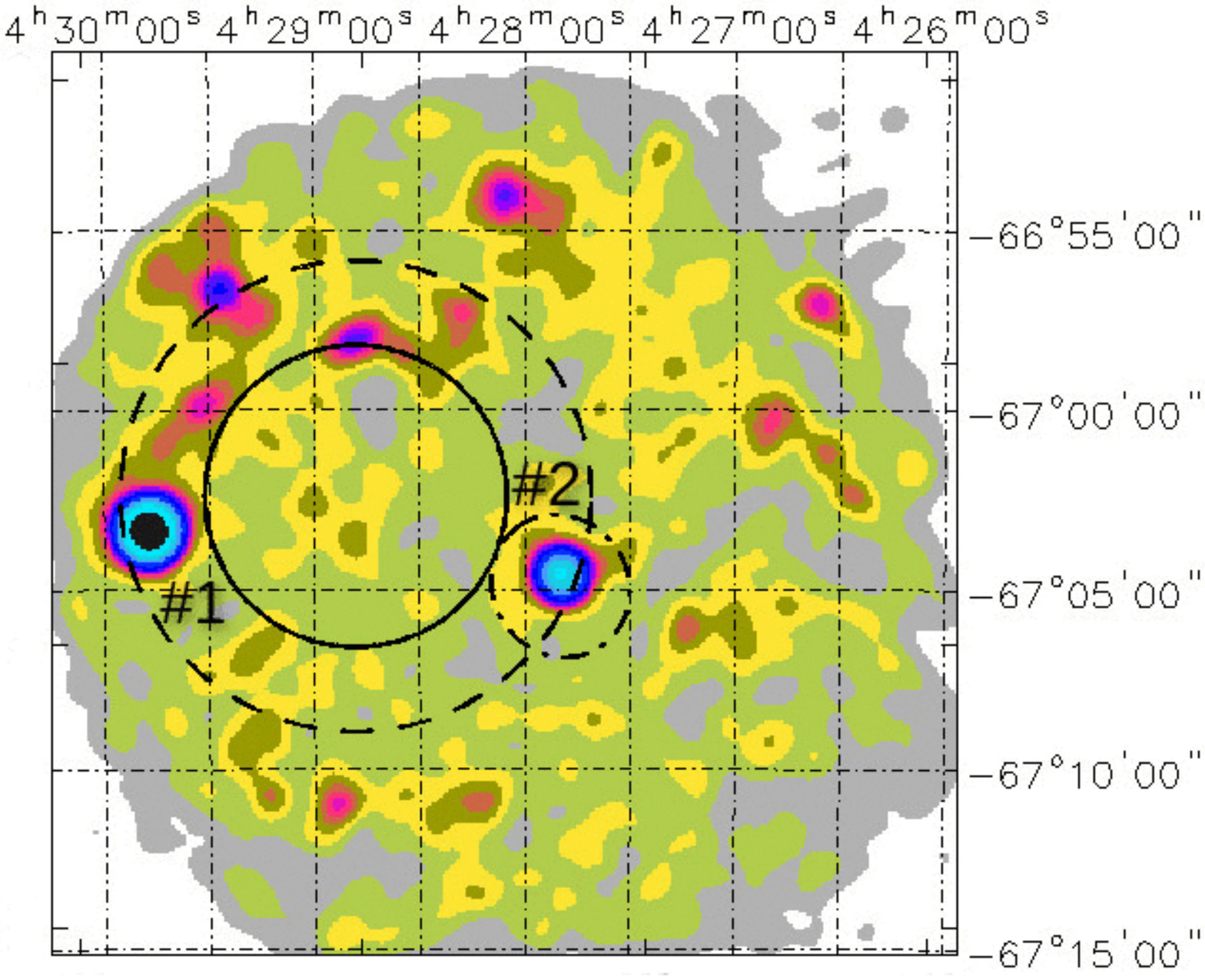}
\includegraphics[width=\textwidth,width=8cm]{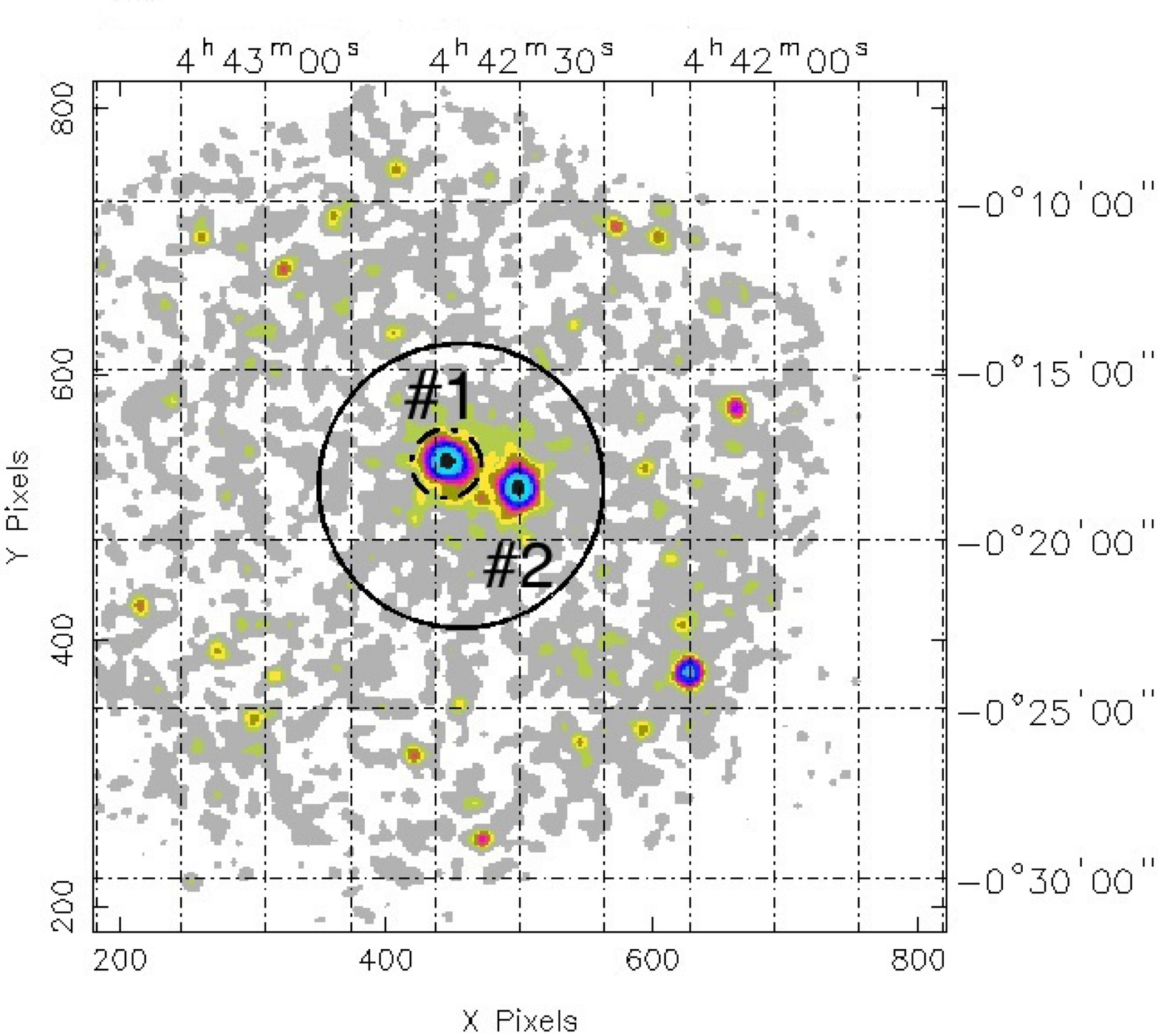}
\caption{{\bf Left}: Swift/XRT 0.3-10 keV image of IGR J04288-6702 also SWIFT J0428.2-6704. The two X-ray counterparts are inside the 99\% IBIS positional uncertainty (dashed circle). Source \#1 is the the Seyfert 1.5 LEDA 299570 at z=0.065, while source \#2 is SXPS J042749.2-670434, the eclipsing low-mass X-ray binary also detected in the Fermi/LAT catalogue (error ellipse). {\bf Right:} 0.3-10 keV band image of PKS 0441-0017 (source \#1) and 1RXS J044229.8-001823 (source \#2), both well inside the 90\% IBIS error circle. The small dashed-dotted ellipse corresponds to the Fermi/LAT positional error of 4FGL J0442.6-0017.}
\label{a1}
\end{figure*}


\begin{figure*}
\includegraphics[width=\textwidth,width=8cm]{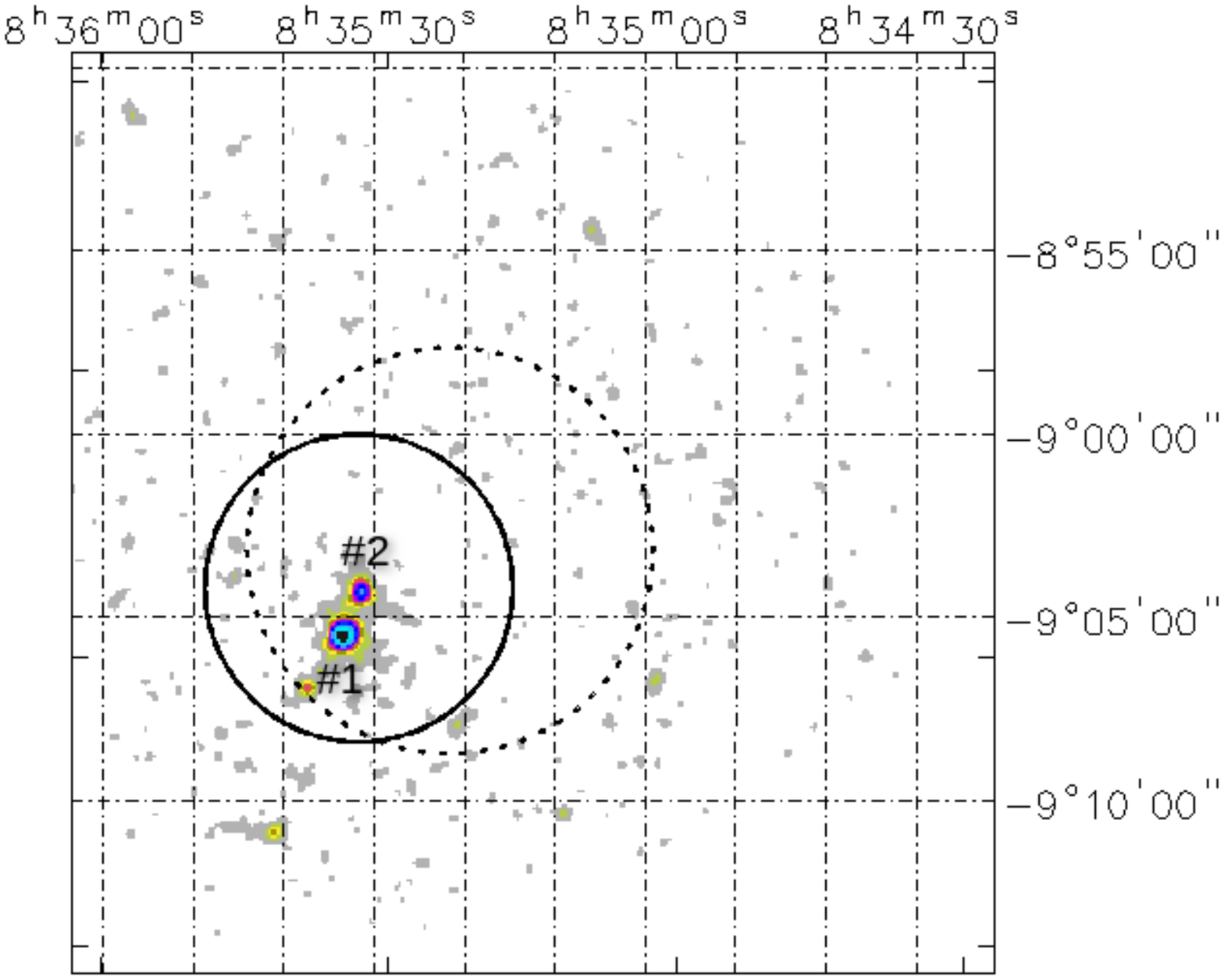}
\includegraphics[width=\textwidth,width=8cm]{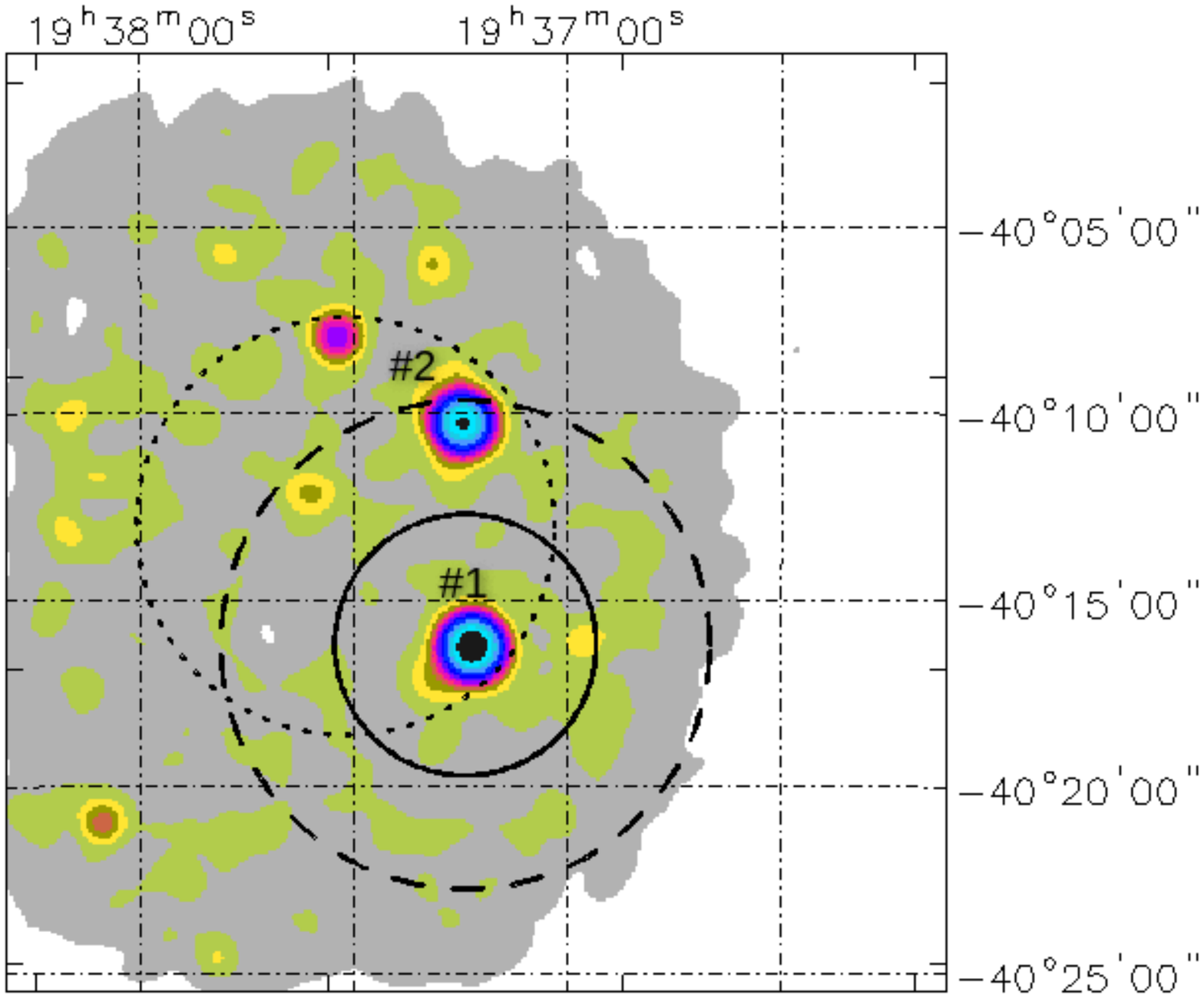}
\caption{{\bf Left}: Swift/XRT 0.3-10 keV image of SWIFT J0835.5-0902. Two X-ray counterparts are well inside the IBIS (continuous circle) and BAT (dotted circle) error circles. Source \#1 is the AGN candidate 2MASX J08353333-0905302, while source \#2 is associated with the variable star ASAS J083531-0904.1. {\bf Right:} 0.3-10 keV image of SWIFT J1937.5-4021. Source \#1, which lies inside the IBIS (continuous circle) and BAT (dotted circle) 90\% positional uncertainty, is the Seyfert 2 LEDA 5882882. Source \#2, which is well inside the BAT error circle, is instead located inside the 99\% IBIS positional uncertainty (dashed circle) and it is associated with 1RXS J193716.1-401026.}
\label{a2}
\end{figure*}


\begin{figure}
\includegraphics[width=\textwidth,width=8cm]{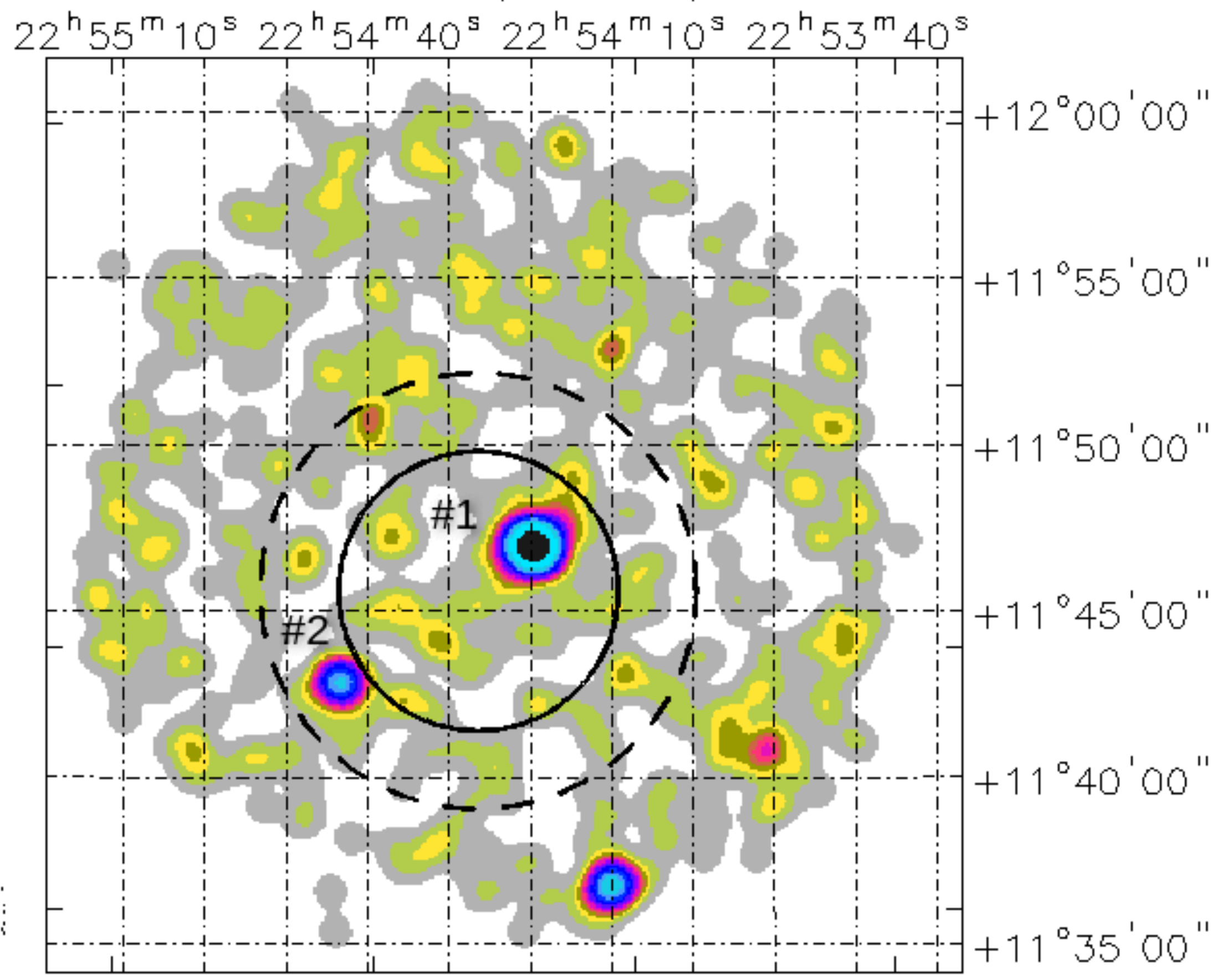}
\caption{Swift/XRT 0.3-10 keV image of AX J2254.3+1146, also called SWIFT J2254.2+1147. Two likely counterparts are detected in X-rays. Source \#1, which falls within the 90\% IBIS positional uncertainty (filled circle) is the Seyfert 2 UGC 12237, while source \#2, which is detected within the 99\% IBIS error circle (dashed circle) is the Seyfert 1 UGC 12243.}
\label{a3}
\end{figure}


\section{Radio images of resolved AGN candidates}
\begin{figure*}
\includegraphics[width=8.3cm]{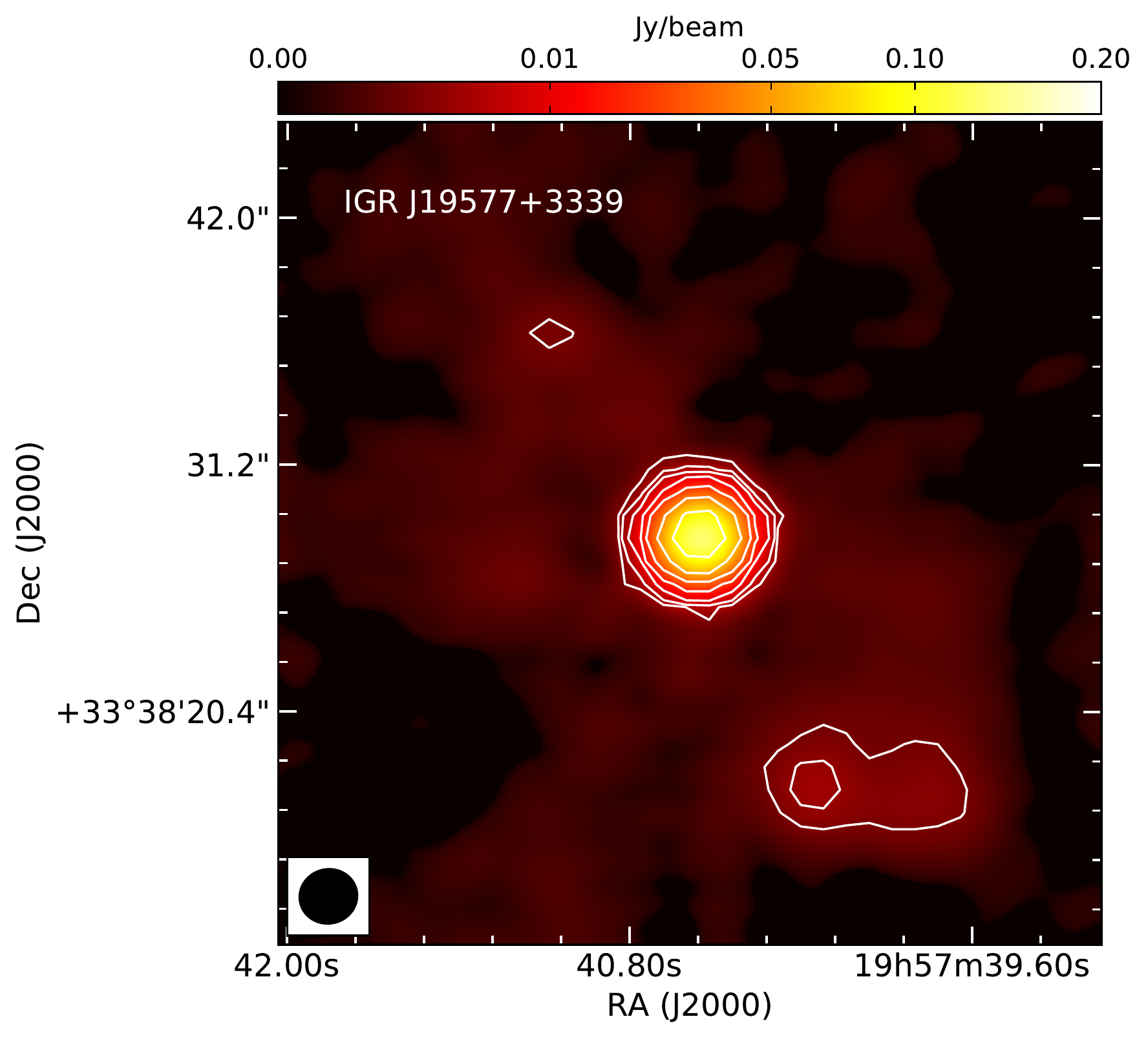}
\includegraphics[width=8cm]{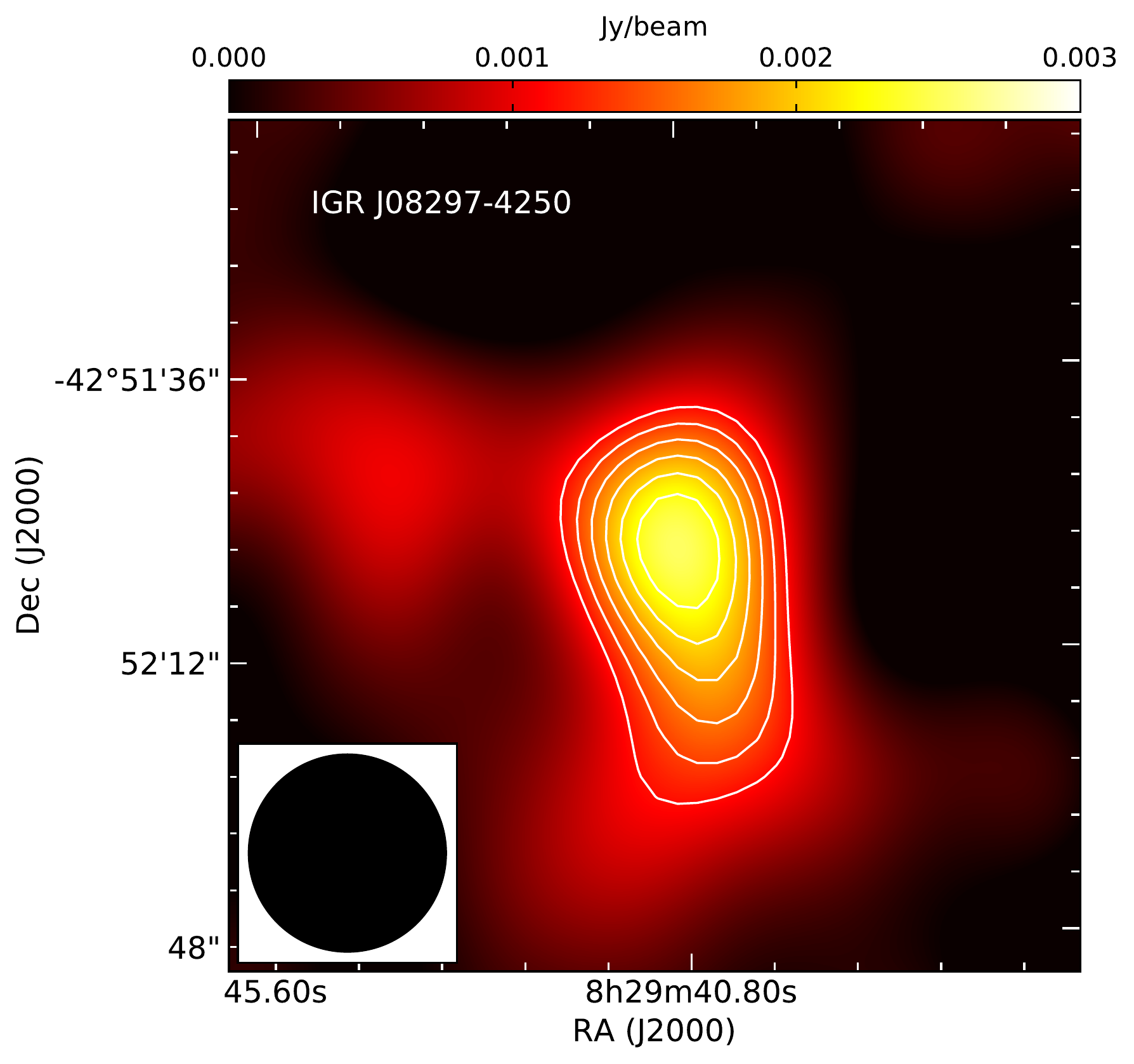}
\caption{Left panel: VLASS image of source IGR\,J19577+3339, contours are 3$\times$RMS$\times$(1,2,4,8,16,32,64). Right panel: RACS image of source IGR\,J08297-4250, contours are RMS$\times$(5,6,7,8,9,10). The angular resolution is shown in the lower-left corner.}
\label{r1}
\end{figure*}

\begin{figure*}
\includegraphics[width=\textwidth,width=8.6cm]{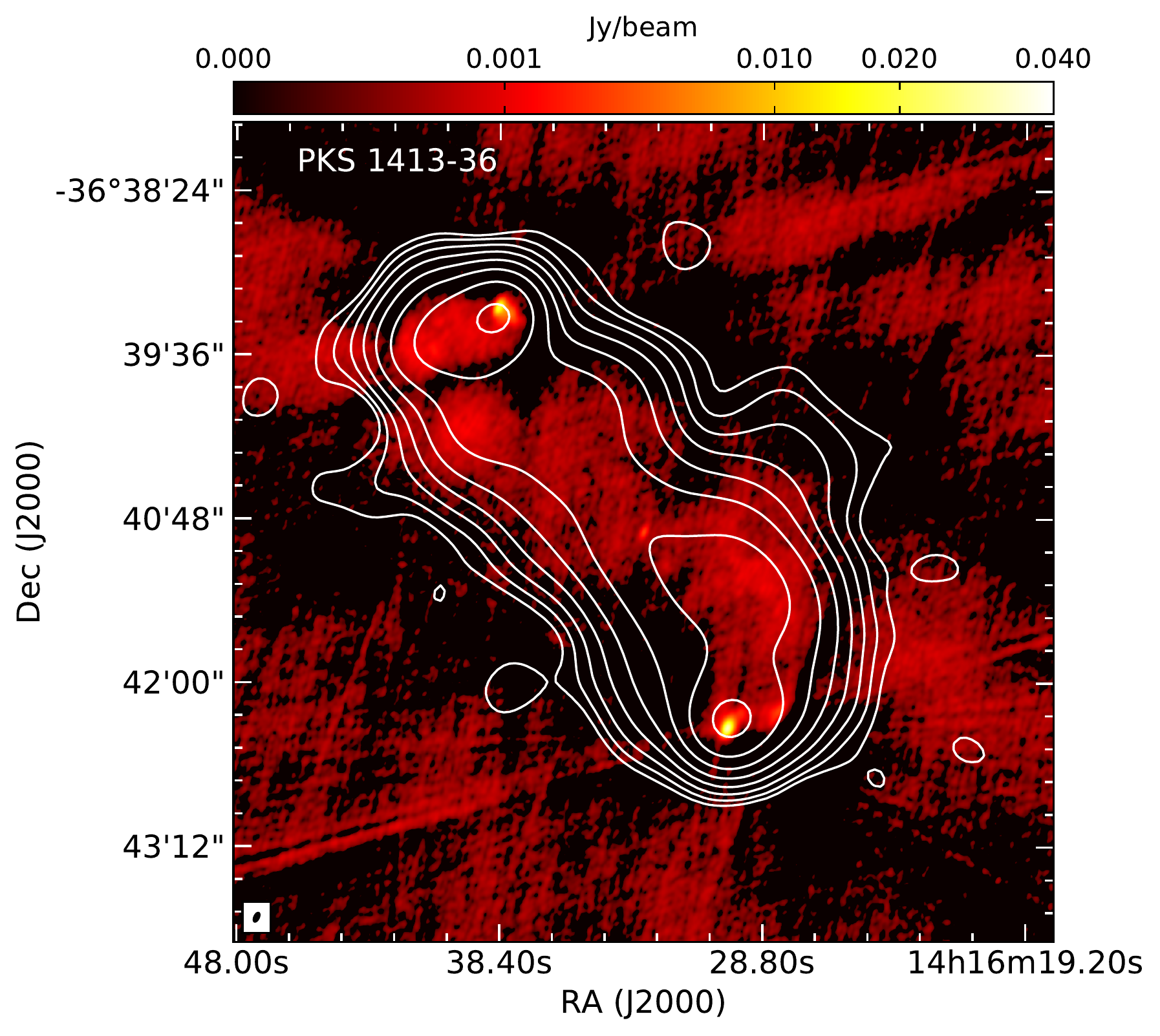}
\includegraphics[width=\textwidth,width=8cm]{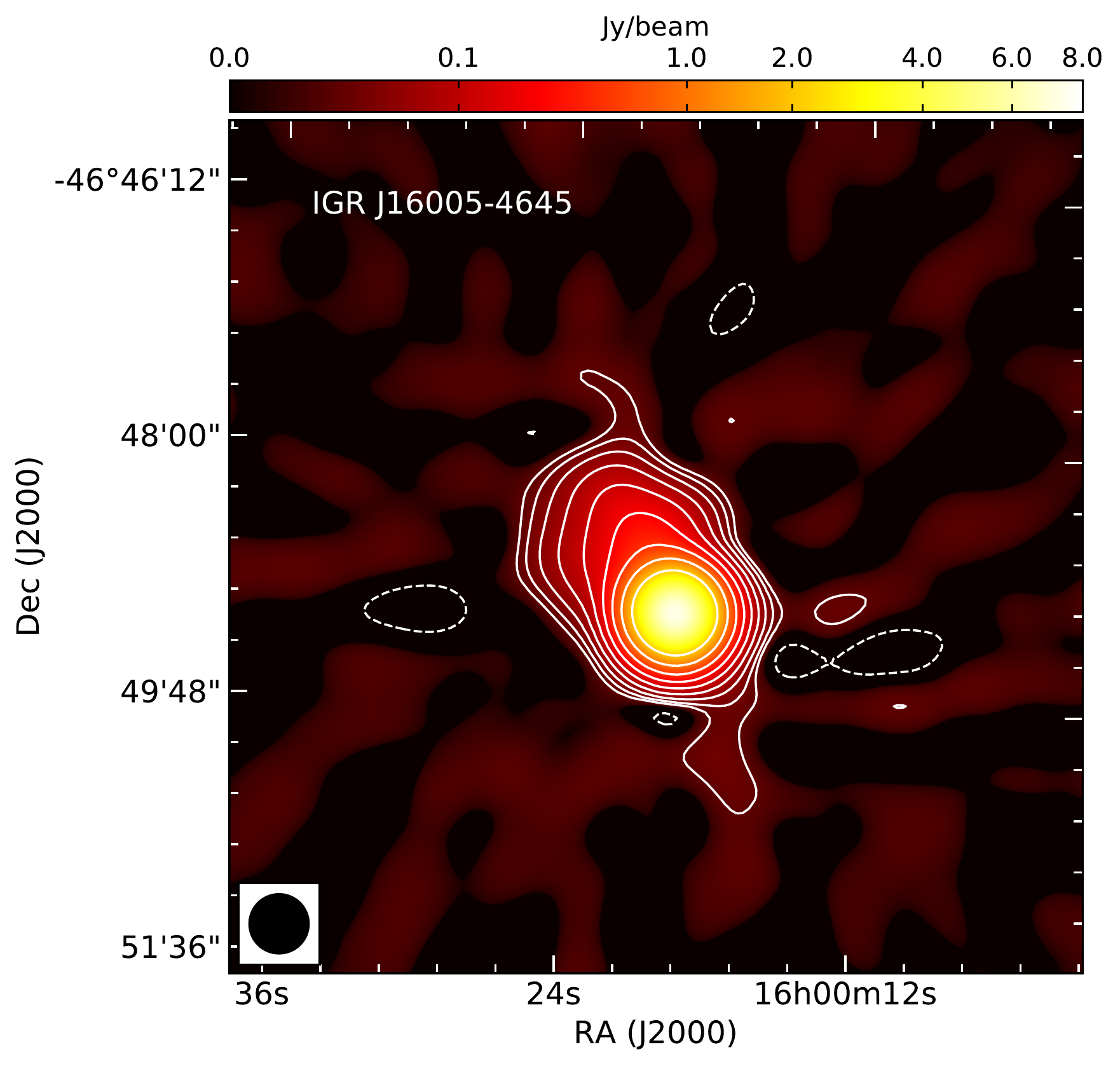}
\caption{Left panel: VLASS image of source PKS\,1413-36 with overlayed contours from RACS, corresponding to 3$\times$RMS$\times$(1,2,4,8,16,32,64,128). Right panel: RACS image of source IGR\,J160005-4645, contours are 3$\times$RMS$\times$(1,2,4,8,16,32,64,128,256). The angular resolution is shown in the lower-left corner.}
\label{r2}
\end{figure*}

The radio images of the resolved sources discussed in section 4.1 are shown in figure \ref{r1} and \ref{r2}. For three of them, only VLASS (IGR J19577+3339) or RACS (IGR J08297-4250 and IGR J16005-4645) images are available, while for the radio galaxy PKS 1413-36 both VLASS and RACS images could be retrieved. VLASS, with a resolution of $\sim$2.5 arcsec and an average RMS of $\sim$0.120 mJy/beam, is more sensitive to compact features ($<$1 arcmin), while RACS, at a resolution of 25 arcsec and average RMS of $\sim$0.25 mJy/beam, is able to recover extended emission up to $\sim$1 deg.

\label{lastpage}
\end{document}